\documentclass[aps,pra,superscriptaddress,twocolumn,10pt]{revtex4-2}

\usepackage{amsmath}
\usepackage{amssymb}
\usepackage{amsthm}
\usepackage{bbm}
\usepackage{color}
\usepackage[T1]{fontenc}
\usepackage{float}
\usepackage{graphicx}
\usepackage{hyperref}
\usepackage{orcidlink}
\usepackage{physics}
\usepackage{times}
\usepackage{txfonts}
\usepackage[utf8]{inputenc}
\usepackage{xcolor}
\usepackage[normalem]{ulem}
\usepackage{lipsum}
\usepackage{tikz}
\usepackage{ulem}

\hypersetup{
    colorlinks=true,linkcolor=blue,citecolor=blue,
    filecolor=blue,urlcolor=blue,breaklinks=true
}

\theoremstyle{definition}

\def\be{\begin{equation}}
\def\ee{\end{equation}}

\def\bc{\begin{center}}
\def\ec{\end{center}}
\def\bal{\begin{align}}
\def\eal{\end{align}}

\begin{document}

\title{Revisiting the Jaynes–Cummings model with time-dependent coupling}

\author{Thiago T. Tsutsui\orcidlink{0009-0001-1654-0330}}
\thanks{Corresponding author:%
\href{mailto:takajitsutsui@gmail.com}{takajitsutsui@gmail.com}}
\affiliation{
 Programa de Pós-Graduação em Ciências/Física,
 Universidade Estadual de Ponta Grossa,
 84030-900 Ponta Grossa, Paraná, Brazil
}

\author{Danilo Cius\orcidlink{0000-0002-4177-1237}}
\email{danilocius@gmail.com}
\affiliation{
Instituto de Física,
Universidade de São Paulo,
05508-090 São Paulo, São Paulo, Brazil
}

\author{Antonio Vidiella-Barranco\orcidlink{0000-0002-6918-8764}}
\email{vidiella@ifi.unicamp.br}
\affiliation{
Instituto de Física Gleb Wataghin,
Universidade Estadual de Campinas,
13083-859 Campinas, São Paulo, Brazil}

\author{Antonio S. M. de Castro\orcidlink{0000-0002-1521-9342}}
\email{asmcastro@uepg.br}
\affiliation{
  Programa de Pós-Graduação em Ciências/Física,
  Universidade Estadual de Ponta Grossa,
  84030-900 Ponta Grossa, Paraná, Brazil
}
\affiliation{
  Departamento de Física, Universidade Estadual de Ponta Grossa
  84030-900 Ponta Grossa, Paraná, Brazil
}

\author{Fabiano M. Andrade\orcidlink{0000-0001-5383-6168}}
\thanks{Corresponding author:%
\href{mailto:fmandrade@uepg.br}{fmandrade@uepg.br}}
\affiliation{
  Programa de Pós-Graduação em Ciências/Física,
  Universidade Estadual de Ponta Grossa,
  84030-900 Ponta Grossa, Paraná, Brazil
}
\affiliation{
  Departamento de Matemática e Estatística,
  Universidade Estadual de Ponta Grossa,
  84030-900 Ponta Grossa, Paraná, Brasil
}

\date{\today}

\begin{abstract}
The Jaynes–Cummings (JC) model stands as a fully quantized, fundamental framework for exploring light–matter interactions, a timely reflection on a century of quantum theory. The time-dependent Jaynes–Cummings (TDJC) model introduces temporal variations in certain parameters, which often require the use of numerical methods. However, under the resonance condition, exact solutions can be obtained, offering insight into a variety of physical scenarios. In this work, we study the resonant TDJC model considering different modulations of the atom–field coupling. The model is presented and an analytical solution derived in a didactic way, allowing us to examine how time-dependent couplings affect atomic population inversion and atom–field entanglement. We also consider an atom traversing a partially cooled cavity, which induces periodicity and reveals the combined effects of atomic motion and thermal fluctuations. The Bloch vector is used to analyze the dynamics of the system, including the atomic state purity, and reveals phenomena such as atomic dipole alignment with the field due to the oscillating coupling, as well as atomic population trapping, which arises by increasing the initial mean thermal photon number.\\

\noindent doi: \href{https://doi.org/10.1007/s13538-025-01949-w}
{10.1007/s13538-025-01949-w}
\end{abstract}

\maketitle

\section{Introduction}
\label{sec:introduction}

In 2025, we mark the centenary of the inception of quantum mechanics, a
revolutionary theory that continues to drive progress through the
ongoing development of quantum technologies.
The birth of quantum theory is deeply connected to the investigation of
electromagnetic (EM) radiation and atoms, which remain central to both
foundational investigations and emerging applications.
More recently, the development of the laser, a coherent light source,
gave rise to the subfield of nonrelativistic quantum electrodynamics
theory known as quantum optics, which basically explores the nature and
effects of quantized light.
In this context, the most elementary quantum model of light–matter
interaction consists of a highly idealized atom (modeled as a two-level
system) coupled to a single mode of the quantized EM field.
Under the rotating wave approximation (RWA), this model, known as the
Jaynes–Cummings (JC) model \cite{JAYNES1963,Larson2024}, remarkably
admits an exact analytical solution, and despite the simplifying
assumptions underlying its construction \cite{DeBernardis2024}, the JC
model provides deep insights into several fundamental aspects of
light–matter interaction.
Originally developed to compare semiclassical and quantum approaches to
spontaneous emission \cite{Cummings2013}, the JC model has since evolved
to uncover nonclassical phenomena such as experimentally observed
collapses \cite{Cummings1965} and revivals of the Rabi oscillations (RO)
\cite{Meystre1975,Eberly1980,MESCHEDE1985,REMPE1987,Brune1996},
squeezing \cite{Kuklinski1988,Hillery1989}, sub-Poissonian statistics
\cite{Hillery1987,Chumakov1993}, atom-field entanglement
\cite{Aravind1984,Phoenix1988,Phoenix1991}, and the generation of
Schrödinger cat states \cite{Guo1996,Gerry1997}.
Furthermore, the core of the model is a two-level system (TLS) coupled
to a harmonic oscillator, which allows it to be mapped onto a trapped ion
\cite{Blockley1992,Vogel1995,Pedernales2015}
and relativistic systems \cite{BERMUDEZ2007}.
Regarding applications, the JC framework is used in several quantum
information processing schemes \cite{LARSON2021,Meher2022}.
There are also generalizations of the model involving additional field
modes and atomic levels \cite{Lai1991,Wu1997},
multiple two-level atoms \cite{Dicke1954,Tavis1968},
quantum deformations \cite{Chaichian1990,Dehghani2016,UHDRE2022},
Kerr-like media \cite{Buzek1990,Buzek1991} and external pumpings  \cite{Bocanegra-Garay2024,Vidiella-Barranco2025}
Besides, the model naturally extends to the description of open quantum
systems, where interaction with the environment leads to dissipation and
decoherence \cite{Puri1986,Barnett1986,Scala2007}.

In 1996, the collapses and revivals of the RO were observed in a
remarkable cavity  quantum electrodynamics experiment by Haroche’s group
\cite{Brune1996}.
Using Rydberg atoms as a TLS \cite{Nussenzveig1993,Brune1994}, this
achievement provided a direct experimental verification of the quantum
nature of the EM field and formed part of the body of work that was
later recognized with the 2012 Nobel Prize in Physics awarded to Serge
Haroche \cite{Haroche2013} jointly with David Wineland because of his
work on trapped ion systems \cite{Meekhof1996,Leibfried2003}.
The JC model's pivotal role in this achievement highlights its
fundamental importance, a fact underscored by the exponential growth of
interest in the system since then \cite{Larson2024}.

A particularly interesting generalization is obtained when the model
parameters are allowed to vary in time, giving rise to the
time-dependent JC (TDJC) model \cite{DeCastro2023,LARSON2021}.
This may account for atomic motion \cite{Schlicher1989,Fang1998},
transient effects in the cavity \cite{Prants1992, Dasgupta1999},
or the incorporation of stochastic aspects \cite{Joshi1995}.
However, unlike the standard JC model, analytical methods are not always
applicable.
Notable exceptions are the Nikitin \cite{Prants1992}, Rosen-Zener
\cite{Dasgupta1999}, Landau-Zener \cite{Larson2003,Keeling2008} and
Demkov–Kunike \cite{Larson2003} models.
Remarkably, under resonance conditions, it is possible to obtain
exact solutions considering a time-dependent atom-field coupling
\cite{Schlicher1989,Joshi1993,Fang1998}.
From an experimental point of view, Ref. \cite{Maldonado-Mundo2012}
suggests that a time-dependent coupling may be achievable in cavity
quantum electrodynamics through variations in the atom's position or by
changing an external electric field.

The objective of this work is twofold: we first present a didactic
overview of the solution of the resonant TDJC model in terms of the
evolution operator.
We discuss two specific cases of time-dependent atom–field coupling:
(i) linear \cite{Joshi1993} and (ii) hyperbolic secant
\cite{Dasgupta1999}.
The latter is particularly desirable in quantum control \cite{Dong2010},
where manipulating the TLS is one potential objective
\cite{Nielsen2010,Hernandez-Sanchez2024}.
We focus on the dynamics of the atomic population inversion and the von
Neumann entropy, examining the impact of the time dependence on both
quantities.
The subsystems are assumed to be initially prepared in pure states, in
which case the entropy directly reflects the atom–field entanglement.
In the second part of the paper, we consider the atom moving inside the
cavity, which leads to a sinusoidal time dependence for the atom–field
coupling. Related scenarios have been explored in the literature,
particularly in studies of entanglement \cite{Bose2001,Yan2009},
self-induced transparency effects \cite{Schlicher1989}, and the role of
atomic coherence in the JC model with motion \cite{Joshi2004}.
We also assume a different set of initial conditions, namely, the field in
a thermal state and the atom prepared either in the excited state or in
a superposition of atomic ground and excited states (eigenstate of
$\hat{\sigma}_x$).
The JC model with a partially cooled cavity is relevant, for example, for
the study of entanglement \cite{Azuma2008,Yan2009}, and is also
experimentally accessible \cite{REMPE1987}.
We analyze the atomic evolution from a different perspective, focusing
on the dynamics of the Bloch vector, and compare our results to the
established findings
\cite{Arancibia-Bulnes1993,Azuma2008,Azuma2014,Azuma2015} within the
context of constant coupling.
We find that the behavior of the quantities studied, basically the
components and modulus of the Bloch vector, is generally dominated by
periodic coupling, even in the presence of thermal noise, in
contrast to the case of constant coupling \cite{Azuma2008}.
We note that a periodic evolution of both the field entropy and the
atomic inversion, induced by atomic motion and in the absence of thermal
noise, was reported in \cite{Fang1998}.
Other effects may arise, manifesting in the Bloch vector dynamics.
Two control parameters in the system, $p$, the number of
half-wavelengths of the standing-wave cavity, and $\zeta_3$,
proportional to the velocity of the moving atom, have similar effects.
For instance, smaller values of $p$ and $\zeta_3$ lead to a dynamical
evolution with longer periods, more closely resembling the behavior of
the constant-coupling case.
Interestingly, increasing the thermal fluctuations, which corresponds to
a larger mean photon number in the initial field state, leads to a form
of atomic population trapping \cite{Arancibia-Bulnes1993}, while the
periodicity of the Bloch vector dynamics tends to persist despite the
added noise.

We organize the paper as follows.
In Sec. \ref{sec:jc_model}, we provide an overview of the standard JC
model, deriving the time-evolution operator and highlighting key aspects
of the dynamics.
Subsequently, in Sec. \ref{sec:time_dep_JC_model}, we examine the
effects of a time-dependent coupling parameter on both the population
inversion and atom-field entanglement, presenting a general
time-evolution operator for this case and analyzing two types of
modulations.
Having established the appropriate formalism, we consider the case of an
atom moving across a partially cooled cavity, focusing on the combined
impact of sinusoidal coupling and thermal fluctuations on the Bloch
vector dynamics.
Finally, our conclusions are summarized in Sec. \ref{sec:conc}.

\section{Jaynes-Cummings model}
\label{sec:jc_model}

\begin{figure}[t!]
\centering
\includegraphics[width=1\linewidth]{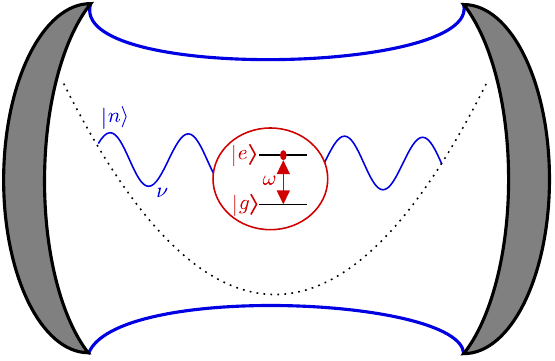}
  \caption{
   A schematic description of the JC model.
   The atom is represented as a TLS, namely $|e\rangle$ and $|g\rangle$,
   with a transition frequency $\omega$, while the cavity mode is
   modeled as a simple harmonic oscillator with frequency $\nu$ and
   characterized by the Fock states $|n\rangle $, $n=0,1,2,\hdots$.
 }
 \label{fig:jc_scheme}
\end{figure}

The JC model is the simplest fully quantum description of light-matter
interaction \cite{Larson2007}.
In the JC framework, matter is represented by one atom, associated with
a single atomic transition, while light is embodied as a quantized mode
of the EM field.
To address this problem, we consider that the Hamiltonian of the system,
$\hat{H}$,  consists of three parts:
the atomic energy $\hat{H}_A$, the EM energy $\hat{H}_F$, and the
interaction term $\hat{H}_I$ \cite{Scully1997,GERRY2005}.
The atom is modelled as a TLS, where $\ket{e}$ and $\ket{g}$ denote the
excited and ground states, respectively.
The field mode, on the other hand, is treated as a simple harmonic
oscillator, with the Fock states $\ket{n}$, $n = 0, 1, 2, \dots$,
serving as a basis.
From the aforementioned concepts and employing the dipole approximation,
we derive the quantum Rabi model \cite{Braak2016}.
Further applying the RWA, we obtain the JC model, whose Hamiltonian is
governed by (hereafter we assume $\hbar=1$)
\begin{equation}
  \label{eq:JC_Hamiltonian_full}
  \hat{H} =
  \frac{1}{2} \omega \hat{\sigma}_z + \nu \hat{a}^\dagger \hat{a}
  + \lambda (\hat{\sigma}_{+}\hat{a}+\hat{\sigma}_{-}\hat{a}^{\dagger}),
\end{equation}
where $\omega$ and $\nu$ denote the frequencies of the atomic transition
and cavity  mode, respectively, while $\lambda$ is the coupling parameter,
assumed to be real.
In Fig. \ref{fig:jc_scheme}, we schematically represent the JC model.

For the TLS, we consider the Pauli matrices $\hat{\sigma}_k$, $k = x, y,
z$, where the ladder operators $\hat{\sigma}_{\pm} = \hat{\sigma}_x \pm
i\hat{\sigma}_y$ serve as generators of the $\mathfrak{su}(2)$ algebra
\cite{Klimov2009}, satisfying the commutation relation
$[\hat{\sigma}_{+}, \hat{\sigma}_{-}] = 2\hat{\sigma}_z$.
In the atomic basis, we can express the aforementioned operators as
$\hat{\sigma}_+ = |e\rangle \langle g|$,
$\hat{\sigma}_- = |g\rangle \langle e|$
and $\hat{\sigma}_z = |e\rangle \langle e| -|g\rangle \langle g|$.
Conversely, the cavity mode is characterized by the bosonic operators,
elements of the Weyl-Heisenberg algebra \cite{Cantuba2024}, obeying
$[\hat{a},\hat{a}^{\dagger}] =  1$.
Their action characterizes the creation (annihilation) of quanta:
$\hat{a}^\dagger |n\rangle=\sqrt{n+1} |n+1\rangle$
($\hat{a} |n\rangle=\sqrt{n} |n-1\rangle$).

The JC system is a composite system comprising two subsystems, the TLS
and the cavity mode, and is therefore classified as bipartite.
The Hilbert space associated with the JC system, $\mathcal{H}$, is
defined as $\mathcal{H} = \mathcal{H}_A \otimes \mathcal{H}_F$,
where $\mathcal{H}_A$ represents the atomic Hilbert space with dimension
$2$, and $\mathcal{H}_F$ corresponds to the cavity Hilbert space,
which has infinite dimension.
As such, the tensor product is used to denote the composite system
state.
Given an atomic state
$|\psi_A\rangle$, $|\psi_A\rangle \in \mathcal{H}_A$, and a field state
$\ket{\psi_F}$, $|\psi_F\rangle \in \mathcal{H}_F$, the initial state of
the system can be in particular expressed as a product state
$|\Psi\rangle = |\psi_A\rangle \otimes |\psi_F\rangle=|\psi_A,\psi_F\rangle$,
provided these subsystems are initially uncorrelated.
In this context, the bare states of the JC model are defined as $
|e,n\rangle$ and $|g,n+1\rangle$,  which are eigenstates of the
excitation number operator \cite{LARSON2021}
\begin{equation}
  \label{eq:excitation_number}
    \hat{N}_E = \hat{a}^\dagger \hat{a} + \frac{1}{2} \hat{\sigma}_z.
\end{equation}
It is straightforward to verify that $[\hat{H}, \hat{N}_E] = 0$,
confirming that $\hat{N}_E$ is a conserved quantity.
This conservation gives rise to a continuous $U(1)$ symmetry, in
accordance with Noether's theorem \cite{DeBernardis2024}, and ultimately
enables the exact solution of the system’s dynamics.
Employing Eq. \eqref{eq:excitation_number}, we can rewrite the
Hamiltonian as
\begin{equation}
  \hat{H} =  \nu \hat{N}_E + \frac{1}{2} (\omega - \nu) \hat{\sigma}_z
  + \lambda
  \left(
    \hat{\sigma}_{+}\hat{a}+\hat{\sigma}_{-}\hat{a}^{\dagger}
  \right).
\end{equation}
In the interaction picture relative to the term $\nu \hat{N}_E$ and
under the resonance condition ($\omega=\nu$), we obtain the interaction
Hamiltonian
\begin{equation}
  \label{eq:int_hamiltonian_jc}
  \hat{V} =
  \lambda (\hat{\sigma}_{+}\hat{a}
  +\hat{\sigma}_{-}\hat{a}^{\dagger}).
\end{equation}

Choosing a basis that diagonalizes $\hat{N}_E$ results in a
block-diagonal structure in the interaction Hamiltonian $\hat{V}$
\cite{Kasper2020}.
Each block $\hat{V}^{(n)}$, corresponds to a specific photon number $n$
and has a $2 \times 2$ dimension, due to the degeneracy of the
eigenvectors of the excitation number operator, $\ket{e,n}$ and
$\ket{g,n+1}$, coupled by the interaction energy expressed in
Eq. \eqref{eq:int_hamiltonian_jc}.
The ground state $\ket{g,0}$ is the only eigenstate of $\hat{N}_E$ that
is not coupled to another bare state, associated with a $1 \times 1$
block and being an eigenvector of the interaction Hamiltonian.
Therefore, employing the bare state basis $\{|e,n\rangle,
|g,n+1\rangle\}$, we can represent $\hat{V}$ in matrix form as
\begin{equation}
\hat{V}=\left[\begin{array}{cccc}
0 & 0 & 0 & \cdots \\
0 & \hat{V}^{(0)} & 0 & \cdots \\
0 & 0 & \hat{V}^{(1)} & \cdots \\
\vdots & \vdots & \vdots & \ddots
\end{array}\right],
\end{equation}
where
\begin{equation}
\hat{V}^{(n)}=\lambda \sqrt{n+1} \left[\begin{array}{ll}
0 & 1 \\
1 & 0
\end{array}\right].
\end{equation}
In the matrix, it is possible to observe how the Hamiltonian
$\hat{V}^{(n)}$ couples the states within a single $2 \times 2$
subspace, with the emission and absorption processes described by the
off-diagonal elements:
$\langle e,n | \hat{V} | g,n+1 \rangle$ corresponds to the transition
$\ket{g,n+1} \rightarrow \ket{e,n}$, while
$\langle g,n+1 | \hat{V} | e,n \rangle$ describes
$\ket{e,n} \rightarrow \ket{g,n+1}$ \cite{Bina2012}.

There are several approaches to studying the dynamics of the JC model
\cite{Juarez-Amaro2015}.
In this context, we focus on the time-evolution operator $\hat{U}(t)$,
primarily because it provides a smooth transition from the standard to
the TDJC model.
Considering the interaction Hamiltonian,
Eq. \eqref{eq:int_hamiltonian_jc}, the time-evolution operator is
defined as \cite{SAKURAI2020}
\begin{equation} \label{eq:time_ev_operator}
    \hat{U}(t) = e^{-i \hat{V} t}.
\end{equation}
We evaluate the exponential employing the series expansion
\begin{equation} \label{eq:expansion_time_ev_op}
    \hat{U}(t) = \sum_{\ell=0}^{\infty} \frac{(-i \hat{V} t)^\ell}{\ell!}.
\end{equation}
The terms of the expansion can be further decomposed by examining their
action on  $|e,n\rangle$.
Inductively:
\begin{align}
  (\hat{V})^{2\ell} \ket{e,n}  = {}
  &( \lambda)^{2\ell} (n+1)^{\ell} \ket{e,n}\quad \mbox{(even)}, \nonumber\\
  (\hat{V})^{2\ell+1} \ket{e,n} = {}
  & ( \lambda)^{2\ell+1} (n+1)^{\ell+1/2} \ket{g,n+1} \quad \mbox{(odd)}.
\end{align}
The action of $(\hat{V})^{2\ell}$ leaves the state unchanged, apart from a multiplicative constant.
In contrast, the action of $(\hat{V})^{2\ell+1}$, in addition to the
constant, results in the inversion of $|e,n\rangle$.
We can extend this analysis to the action on the state $\ket{g,n+1}$,
and, subsequently, separate the even and odd terms in the expansion of
$\hat{V}$  \cite{Phoenix1988}:
\begin{align}
  \label{eq:even_odd_exp}
  \left(\hat{\sigma}_{+} \hat{a}+\hat{\sigma}_{-} \hat{a}^{\dagger}
  \right)^{2 \ell}  = {}
  &
    \left(\hat{a} \hat{a}^{\dagger}\right)^{\ell}|e\rangle \langle e|
    +\left(\hat{a}^{\dagger} a\right)^{\ell} |g\rangle \langle g|,
    \nonumber \\
  \left(\hat{\sigma}_{+} \hat{a}+\hat{\sigma}_{-} \hat{a}^{\dagger}
  \right)^{2 \ell+1} ={}
  &
    \left(\hat{a} \hat{a}^{\dagger}\right)^{\ell} \hat{a}
    |e\rangle \langle g|
    +\hat{a}^{\dagger}\left(\hat{a} \hat{a}^{\dagger}\right)^{\ell}
    |g\rangle \langle e|.
\end{align}
Employing Eq. \eqref{eq:even_odd_exp} and algebraic manipulations, we
can express $\hat{U}(t)$, in the more convenient form of
\begin{align}
  \hat{U}(t) = {}
  &
    \sum_{\ell =0}^{\infty}
    \left\{
    \frac{(-i \lambda t)^{2 \ell}}{(2 \ell)!}\left[\left(\sqrt{\hat{a}
    \hat{a}^{\dagger}}\right)^{2 \ell}|e\rangle \langle e|
    +
    \left(\sqrt{\hat{a}^{\dagger} \hat{a}}\right)^{2 \ell} |g\rangle
    \langle g| \right] \right.
    \nonumber \\
  &
    +\left. \frac{(-i \lambda t)^{2 \ell+1}}{(2
    \ell+1)!}\left[\frac{\left(\sqrt{\hat{a}
    \hat{a}^{\dagger}}\right)^{2 \ell+1}}{\sqrt{\hat{a}
    \hat{a}^{\dagger}}} \hat{a} |e\rangle \langle g| +\hat{a}^{\dagger}
    \frac
    {\left(\sqrt{\hat{a} \hat{a}^{\dagger}}\right)^{2 \ell+1}}
    {\sqrt{a \hat{a}^{\dagger}}}|g\rangle \langle e|\right]\right\}.
\end{align}
By using the series definition of the sine and cosine functions
\cite{Abramowitz1972}, we can express the operator with the closed
expression \cite{LARSON2021}
\begin{align}
  \label{eq:time_ev_op_explicit_all}
  \hat{U}(t)= {}
  & \cos (\lambda t \sqrt{\hat{a}^\dagger \hat{a}+1})
    |e\rangle \langle e| +\cos (\lambda t
    \sqrt{\hat{a}^\dagger \hat{a}}) |g\rangle \langle g|
    \nonumber \\
  &
    -i
    \frac{\sin (\lambda t \sqrt{\hat{a}^\dagger \hat{a}+1})}
    {\sqrt{\hat{a}^\dagger \hat{a}+1}} \hat{a} |e\rangle \langle g|
    -i \hat{a}^{\dagger} \frac{\sin (\lambda t \sqrt{\hat{a}^\dagger \hat{a}+1}))}{\sqrt{\hat{a}^\dagger \hat{a} +1}}|g\rangle \langle e|.
\end{align}

Equipped with the time evolution operator $\hat{U}(t)$, we can analyze
the system's dynamics.
We consider the initial state
\begin{equation} \label{eq:initial_state}
    |\Psi(0)\rangle = |e, \alpha\rangle,
\end{equation}
where $|\alpha\rangle = \sum_{n=0}^\infty C_n |n\rangle$ denotes a
coherent state \cite{Glauber1963c}.
Coherent states play a fundamental role in quantum optics and in the
study of the JC model, as they are minimum-uncertainty states that
closely resemble a classical description of the EM field.
They have a Poissonian photon number distribution and,
unlike nonclassical states such as Fock states, are more readily
accessible in experiments \cite{Zhang2024}.
The expansion coefficients are given by
\begin{equation}  \label{eq:coefficients_coherent}
  C_n = e^{-|\alpha|^2/2} \frac{\alpha^n}{\sqrt{n!}},
\end{equation}
where $|\alpha|^2 = \langle n \rangle$ represents the initial average
photon number in the cavity.

The system evolves according to
$|\Psi(t)\rangle = \hat{U}(t) |\Psi(0)\rangle$, which leads to
\cite{Scully1997,GERRY2005}
\begin{equation}
  \label{eq:coherent_jc_state_t_resssonance}
  \ket{\Psi(t)}  =\sum_{n=0}^\infty \left[ C_{e,n} (t) |e\rangle
    + C_{g,n} (t) |g\rangle \right] |n\rangle,
\end{equation}
where the probability amplitudes are given by
\begin{align}
\label{eq:prob_amps}
  C_{e,n} (t)  = {}
  & C_n \cos\left(2 \lambda \sqrt{n+1} t\right) \nonumber\\
  C_{g,n} (t)  = {}
  & -i C_{n-1} \sin\left(2 \lambda \sqrt{n} t\right).
 \end{align}
Taking the square modulus of Eq. \eqref{eq:prob_amps} and marginalizing
over the Fock states, we obtain the probabilities of finding the TLS in
the excited and ground states, respectively,
\begin{align}
  \label{eq:atomic_probs}
  P_{e} (t)  = {} & \sum_{n=0}^\infty | C_{e,n} (t) |^2,  \nonumber\\
  P_{g} (t)  = {} & \sum_{n=0}^\infty | C_{g,n} (t) |^2.
\end{align}
In this context, a quantity of interest is the expectation value
$W(t) = \langle \hat{\sigma}_z \rangle$, referred to as the population
inversion, which can be written as
 \begin{equation}
   W(t) = P_e (t) - P_g (t).
\end{equation}
This function is experimentally measurable
\cite{LARSON2021,Haroche2006}, often computed within the JC model and
its extensions \cite{Arroyo-Correa1990}.

For the initial state $|\Psi (0) \rangle = |e,\alpha \rangle$, the
population inversion is
\begin{equation}
W(t) = \sum_{n=0}^\infty P_n \cos\left(2 \lambda \sqrt{n+1} t \right),
\end{equation}
where $P_n = |C_n|^2$ if the Poissonian photon number distribution of
the initial coherent field state $|\alpha\rangle$, with $\langle n
\rangle = |\alpha|^2$.
The sum of the trigonometric functions, weighted by $P_n$, results in
the collapses and revivals of the RO
-- a hallmark of the JC model that underscores the quantum signature of
the EM field.
This phenomenon arises as RO dephase over time due to the spread of the
Rabi frequencies caused by the photon number distribution of the
coherent state, leading to their ``collapse''.
Rephasing occurs at later times, causing the oscillations to ``revive'',
with the first revival occurring approximately at
$t_r = 2 \pi \sqrt{\langle n \rangle}/ \lambda$  \cite{LARSON2021}.

Since the state in Eq. \eqref{eq:coherent_jc_state_t_resssonance}
cannot, in general, be written as a product state of the form
$|\Psi\rangle=|\psi_A\rangle\otimes|\psi_F\rangle$, it is typically
non-separable \cite{Nielsen2010}.
This reflects the phenomenon of entanglement, where two or more systems
become so strongly correlated that they can no longer be described
independently.
Mathematically, this arises from the quantum formalism, where the
system's Hilbert space is described using the tensor product of its
subsystems Hilbert spaces.
Consequently, the resulting state vector is generally not a product of
the individual states of the system \cite{Horodecki2009}.
Historically, the investigation of entanglement is deeply intertwined
with the development of quantum mechanics.
The phenomenon was first explicitly discussed by Schrödinger in 1935,
who coined the term entanglement (\emph{verschränkung}) to describe the
peculiar quantum correlations between particles \cite{Schrodinger1935}.
Earlier that same year, Einstein, Podolsky, and Rosen had published
their seminal paper \cite{Einstein1935}, which, while not using the term
entanglement, highlighted the odd nature of quantum correlations and
challenged the completeness of quantum mechanics.
Later, in 1964, Bell formulated his famous theorem, which demonstrated
that the predictions of quantum mechanics for entangled pure states
cannot be explained by any local hidden variable theory, thereby
emphasizing the nonlocal nature of quantum entanglement \cite{Bell1964}.

\begin{figure*}
  \centering
  \includegraphics[width=0.65\linewidth]{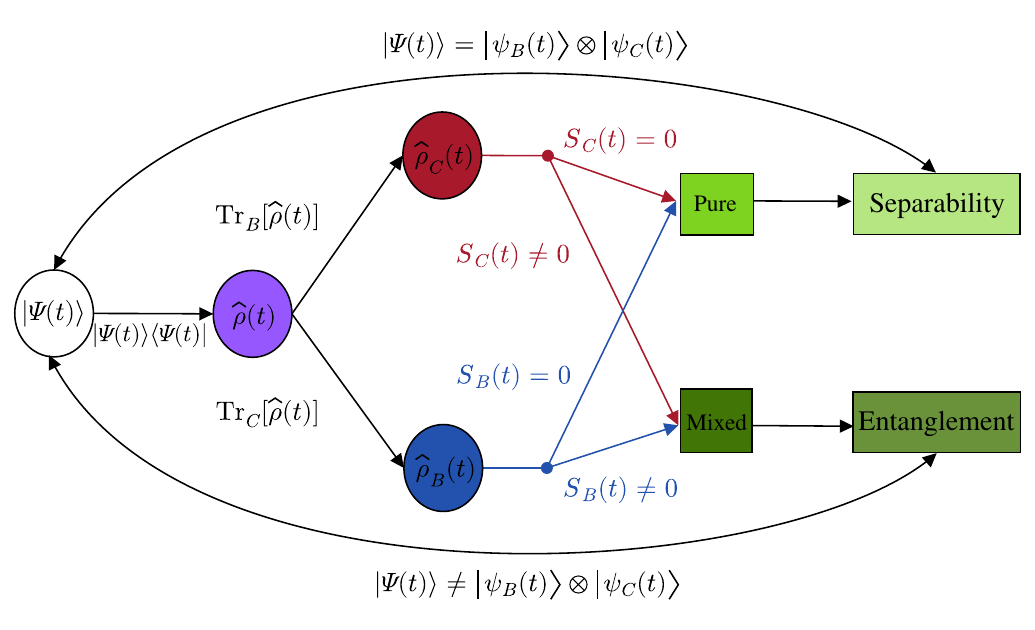}
  \caption{
    Scheme for the relation between the von Neumann entropy and
    entanglement.
    We begin with an initial pure state
    $\ket{\Psi(t)}\in \mathcal{H}_{BC}=\mathcal{H}_B \otimes \mathcal{H}_C$,
    and describe it using the density operator formalism to obtain
    $\hat{\rho}(t) = |\Psi(t)\rangle\langle\Psi(t)|$.
    Sequentially, we arbitrarily trace out the degree of freedom $C$,
    resulting in the reduced density matrix
    $\hat{\rho}_B(t) = \operatorname{Tr}_C\left[\hat{\rho}(t)\right]$.
    We then compute the von Neumann entropy
    $S_B(t) = -\operatorname{Tr}\left[\hat{\rho}_B(t) \log \hat{\rho}_B(t)\right]$
    to quantify the amount of mixedness.
    A zero entropy indicates that the reduced density matrix
    $\hat{\rho}_B(t)$ is pure, and hence the state $|\psi_{BC}\rangle$
    is separable.
    Conversely, a nonzero entropy implies that $\hat{\rho}_B(t)$ is a
    mixed state, signaling that the state $|\Psi(t)\rangle$ is
    entangled.
    In this scenario, the mixedness is proportional to the entanglement,
    as it reflects the uncertainty in the individual subsystem.
  }
  \label{fig:vne}
\end{figure*}

In this context, for pure states in bipartite systems, the von Neumann
entropy \cite{vonNeumann1927b} provides a well-established measure of
entanglement \cite{Phoenix1990,LARSON2021}.
While the von Neumann entropy quantifies the mixedness of a density
matrix in general, when computed for the reduced density matrix of one
subsystem, it effectively measures the degree of entanglement between
the two subsystems.
This occurs because, the more correlated the system is, the less
information is retained in the reduced subsystem after the partial
trace, leading to a more mixed state.

The Schmidt decomposition \cite{Schmidt1907,Ekert1995} is particularly
useful when dealing with two subsystems, one of which has dimension $m$
($m \neq 2$).
It guarantees that, for a bipartite system in a pure global state, there
exists an orthonormal basis
$\{ \ket{b_i(t)}, i = 1, 2, 3, \dots \}$ for subsystem $B$ and
$\{ \ket{c_j(t)}, j = 1, 2, 3, \dots \}$ for subsystem $C$, such that
the state of the system, $\ket{\psi_{BC}} \in \mathcal{H}_{BC}$, can be
written as
\begin{equation}
  \label{eq:schmidt_dec}
  \ket{\psi_{BC}(t)} = \sum_i \mu_i(t) \ket{b_i(t), c_i(t)}.
\end{equation}
In the expression above, the upper limit of the index $i$ is the
\emph{smaller} of the dimensions of the Hilbert spaces involved.
In the case of the JC model, for instance, the Schmidt decomposition
allows the state vector to be expressed in terms of just two terms—even
though the cavity’s Hilbert space is infinite-dimensional.
This greatly facilitates the computation of the von Neumann entropy.

The von Neumann entropy is defined as
$S_j(t)=-\Tr [\hat{\rho}_j (t) \log_2{\hat{\rho}_j(t)}]$,
where $j$ denotes the specific subsystem.
The maximum entropy value is $ \log_2(\dim{\mathcal{H}_i}) $, and base
two is chosen arbitrarily so that the maximum value is $1$ when
evaluating the subsystems of the JC model.
In a basis where $\hat{\rho}_j(t)$ is diagonal with eigenvalues
$\mu_i(t)$, the von Neumann entropy is given by
\begin{equation}
  \label{eq:vne}
  S_j(t) = -\sum_i \mu_i(t) \log_2{\mu_i(t)}.
\end{equation}
Notably, for pure initial states, the entropies of the reduced
subsystems are equal, i.e., $S_B(t) = S_C(t)$ \cite{Araki2002}, where
$j = B, C$ represents generic correlated subsystems \cite{LARSON2021}.
In Fig. \ref{fig:vne}, we provide a schematic representation of the
relationship between the von Neumann entropy and entanglement.

\begin{figure}[h]
  \centering
  \includegraphics[width=1\linewidth]{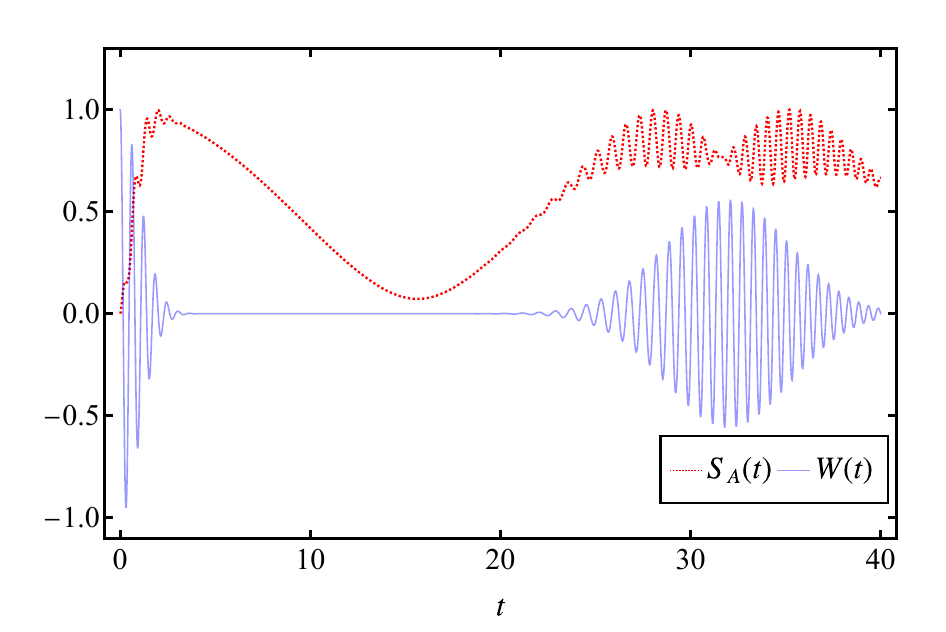}
  \caption{
    The population inversion (solid blue line) and von Neumann entropy
    (dotted red line) as a function of the dimensionless time, when
    considering the initial state
    $|\Psi (0) \rangle = |e,\alpha \rangle$, an average photon number of
    $\langle n \rangle = 25$, and a constant coupling parameter
    $\lambda_0 = 1$.
  }
  \label{fig:vne_const_coup}
\end{figure}

Atom-field entanglement is one of the hallmark nonclassical features of
the JC model \cite{Phoenix1988,Phoenix1990,Phoenix1991}.
Historically, the purity of the state was first studied by Aravind and
Hirschfelder \cite{Aravind1984}.
A more rigorous analysis was conducted by Phoenix and Knight in 1988
\cite{Phoenix1988}, with the term “entanglement” being used for the
first time in 1991 \cite{Phoenix1991} by the same authors.
Since then, due to its relevance in quantum computing applications, the
nonseparability of the state in the JC model has been extensively
investigated
\cite{Boukobza2005,Fasihi2019,Yonac2006,Prants2006,Tan2011,Cheng2018}.
To compute entanglement in the context of the JC model, we begin with
the evolved state described in
Eq. \eqref{eq:coherent_jc_state_t_resssonance} and its corresponding
probability amplitudes, Eq. \eqref{eq:prob_amps}.
By definition, the density matrix of the \emph{system} is
$\hat{\rho} (t)= |\Psi(t)\rangle \langle \Psi(t) |$.
Tracing out the cavity degree of freedom we obtain the reduced density
operator for the atom
\begin{equation} \label{eq:Trace}
  \hat{\rho}_A (t)= \Tr_F{[\hat{\rho}} (t)] =
  \sum_{n=0}^\infty \langle n | \hat{\rho} (t) | n \rangle,
\end{equation}
matricially represented by
\begin{equation} \label{eq:density_matrix_bare}
 \hat{\rho}_A (t) =   \begin{pmatrix}
   P_e(t) & \xi(t)\\
    \xi^*(t)  &  P_g(t)
\end{pmatrix}.
\end{equation}
The coherence term is given by
$\xi(t) = \sum_{n=0}^{\infty} C^*_{e,n}(t)  C_{g,n} (t)$
and the diagonal elements are the atomic populations,
Eq. \eqref{eq:atomic_probs}.
The instantaneous eigenvalues can be computed from the characteristic
equation $\det[ \hat{\rho}_A (t) - \mu_i (t) \hat{I}] = 0$, and are
given by
\begin{equation}
  \mu_{\pm}(t)
  =\frac{1}{2}\left(1\pm\sqrt{W^2(t)
      +4 \left|\xi(t) \right|^{2}}\right),
\end{equation}
can be used to compute the von Neumann entropy, Eq. \eqref{eq:vne}.
The behavior of the entanglement in this situation is presented in
Fig. \ref{fig:vne_const_coup}, alongside the population inversion.
We observe that the system is nearly separable approximately at the
center of the collapse region, while exhibiting relatively high
entanglement during the revival phase.
In general, the separability of the system can be understood by
expressing the joint atom-field state in the eigenbasis that
diagonalizes
$\hat{\rho}_A(t)$:
\begin{equation} \label{eq:state_schmidt}
  |\Psi (t) \rangle
  = \vartheta_{+}(t) |\psi_{A}^{+} (t),\psi_{F}^{+}(t)\rangle +
    \vartheta_{-}(t) |\psi_{A}^{-}(t),\psi_{F}^{-} (t)\rangle,
\end{equation}
where $|\vartheta_{\pm}(t)|^{2}=\mu_\pm(t)$,
$\{ |\psi_{A}^{i} (t)\rangle, i=+,- \}$ and
$\{ |\psi_{F}^{i} (t)\rangle, i=+,- \}$, represent time-dependent
eigenbases.
For example, the entanglement is nearly zero when $\mu_+(t) \gg
\mu_-(t)$, which happens in the collapse region \cite{Phoenix1990}.
As a consequence, the atom-field state can be written as $  |\Psi (t)
\rangle \approx |\psi_A^+(t),\psi_F^+(t)\rangle$.
We emphasize that Eq. \eqref{eq:state_schmidt} expresses the Schmidt
decomposition of the JC model.

\section{Time-dependent coupling dynamics}
\label{sec:time_dep_JC_model}

As mentioned in Sec. \ref{sec:introduction}, the Hamiltonian of the TDJC
model is a modified version of the JC Hamiltonian,
Eq. \eqref{eq:JC_Hamiltonian_full}, with explicit time dependence in
parameters that are usually fixed.
This approach was first implemented by Schlicher \cite{Schlicher1989},
who considered a sinusoidal coupling parameter and studied transparency
effects.
Solving the TDJC model off-resonance often requires numerical methods,
with analytical solutions being the exception unless additional
approximations are applied
\cite{Larson2003,Dasgupta1999,Prants1992,Keeling2008}.
A recent and compelling development in this area is the utilization of
the symmetry properties of the JC Hamiltonian to analyze the system
\cite{DeCastro2023}.

In this work, we are interested in how time-dependent coupling
parameters affect the on-resonance dynamics of the JC model.
Other works that adopt a similar approach can be found in Refs. \cite{Schlicher1989,Bartzis1992,Joshi1993,Fang1998,Yang2006}.
We remark that stochastic effects can be incorporated using a
time-dependent coupling, as demonstrated in Ref. \cite{Joshi1995}.
Therein, phase fluctuations are modeled via a random telegraph process
to study the effects of decoherence.
In this approach, the oscillations of both the atomic population
inversion and the second-order correlation function of the field exhibit
damping that is inversely proportional to the mean time interval between
stochastic phase changes.
While this is an interesting scenario, here we focus on well-behaved
time-dependent couplings, and a detailed treatment of stochastic
couplings is not addressed within the present analysis.

The change $\lambda \to \lambda(t)$ makes the Hamiltonian, Eq. \eqref{eq:int_hamiltonian_jc}, explicitly time-dependent,
while it still commutes with itself at different instants.
As a result, the time evolution operator is described as
\begin{equation} \label{eq:time_ev_operator_timedep}
  \hat{U}(t) = \exp\left[-i \int^{t}_{0} \hat{V}(t^{\prime})
    \dd t^{\prime}\right],
\end{equation}
instead of the expression in Eq. \eqref{eq:time_ev_operator}
\cite{SAKURAI2020}.
Here, we introduce the coupling-area, defined as
\cite{Schlicher1989,LARSON2021}
\begin{equation} \label{eq:coupling_area}
    A(t)=\int^{t}_{0} \lambda (t^{\prime}) \dd t^{\prime}.
\end{equation}
The algebraic procedures realized in Sec. \ref{sec:jc_model} remain the
same, with the change $\lambda t \to A(t)$ in the final result
\cite{Yan2009}.
Thus
\begin{align}
  \label{eq:time_dep_coup_time_ev_op_explicit_all}
  \hat{U}(t)= {}
  &
    \cos \left[   A(t) \sqrt{\hat{a}^\dagger \hat{a}+1} \right]
    |e\rangle\langle e|
        \nonumber \\
  &
    + \cos \left[  A(t) \sqrt{\hat{a}^\dagger \hat{a}} \right]
    |g\rangle\langle g|
    \nonumber \\
  &
    -i \frac{\sin \left[  A(t)\sqrt{\hat{a}^\dagger
    \hat{a}+1}\right]}{\sqrt{\hat{a}^\dagger
    \hat{a}+1}}\hat{a} |e\rangle\langle g|
    \nonumber\\
  &
    -i  \frac{\sin \left[  A(t) \sqrt{\hat{a}^\dagger
    \hat{a}+1}\right]}{\sqrt{\hat{a}^\dagger \hat{a} +1}}
    \hat{a}^{\dagger}|g\rangle\langle e|.
\end{align}

Fixing the initial state as $|\Psi(0)\rangle=|e,\alpha\rangle$ and
employing the time-evolution operator,
Eq. \eqref{eq:time_dep_coup_time_ev_op_explicit_all},
the probability amplitudes can be generalized for an arbitrary
time-dependent coupling as follows
\begin{equation}
  \begin{aligned}
    \label{eq:coefficients_time_dep_coup_generalized}
        C_{e,n}(t) &= C_n \cos \left[A(t) \sqrt{n+1} \right] , \\
        C_{g,n}(t) &= -i C_{n-1} \sin \left[A(t) \sqrt{n} \right],
    \end{aligned}
\end{equation}
where $C_n$ is governed by Eq. \eqref{eq:coefficients_coherent}.
The same approach as before can be applied to the population inversion,
yielding
\begin{equation}
  \label{eq:generalized_pop_inversion}
    W(t) = \sum_{n=0}^{\infty} P_n \cos \left[2 A(t) \sqrt{n+1} \right].
\end{equation}
Here $P_n=|C_n|^2$ is the photon number probability distribution of the
initial field state.
The reason the JC model retains much of its structure in this
generalization is that it preserves its block structure, even with the
time-dependent coupling parameter \cite{LARSON2021}.
This occurs because the excitation number operator,
Eq. \eqref{eq:excitation_number}, remains a constant of motion, which
can be verified by computing its expectation value.

In the following subsections, we will compute the population inversion
and entanglement for two distinct time dependencies in the coupling
parameter.
The procedure is straightforward:
(i) we first calculate the coupling area, $A(t)$;
(ii) using Eq. \eqref{eq:coefficients_time_dep_coup_generalized}, we
compute the state vector;
(iii) from the state vector, we obtain the density operator;
(iv) from the density operator, we calculate the entropy;
and (v) using  Eq. \eqref{eq:generalized_pop_inversion}, we compute the
population inversion.
Additionally, we note that our presentation of the modulation forms
differs from that in the original references, as we express them in
terms of two constant parameters, $\lambda_0$ and $\zeta$, both having
the same dimension as $\lambda(t)$.

\subsection{Linear}
Joshi and Lawande \cite{Joshi1993} considered the linear coupling
modulation given by
\begin{equation}
  \label{eq:linear_coup}
    \lambda(t)=\lambda_0 \zeta_1 t,
\end{equation}
where $\lambda_0$ and $\zeta_1$ are control parameters that determine
whether the coupling changes suddenly or adiabatically.
According to \cite{LARSON2021}, this modulation can be used to
represent the scenario of a well-localized atom in a spatially varying
mode.
The authors observed that, for an initial coherent field state, this
time-dependent coupling alters the revival time.
More recently, Hu and Tan \cite{Hu2014} studied a double JC model with
linear modulation, focusing on the disappearance of entanglement between
the two atoms.
They suggested that a linear coupling parameter could be realized in
cavity quantum electrodynamics by controlling the intensity of an
external electric field.

\begin{figure}[t!]
  \centering
  \includegraphics[width=1\linewidth]{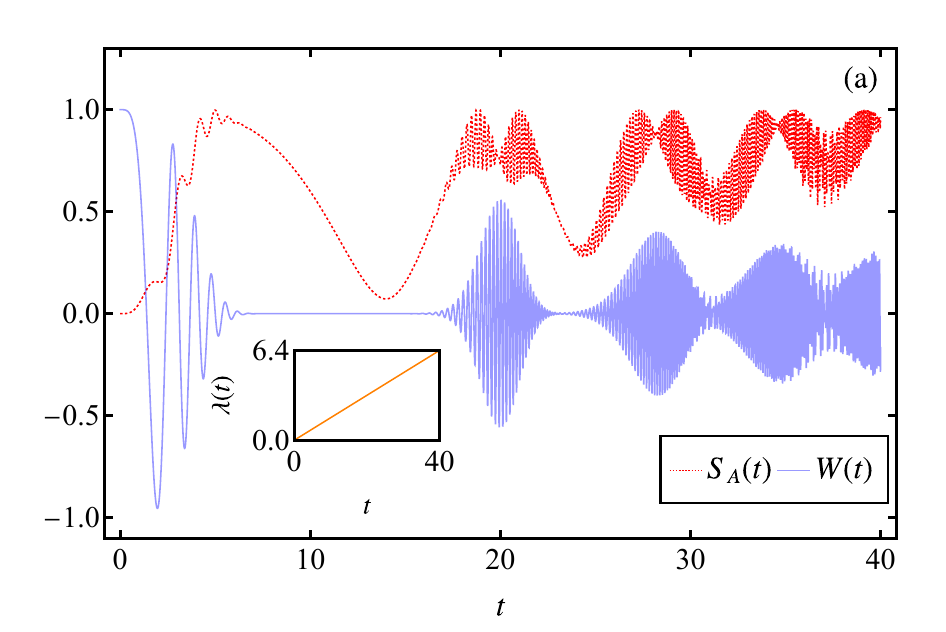}
  \includegraphics[width=1\linewidth]{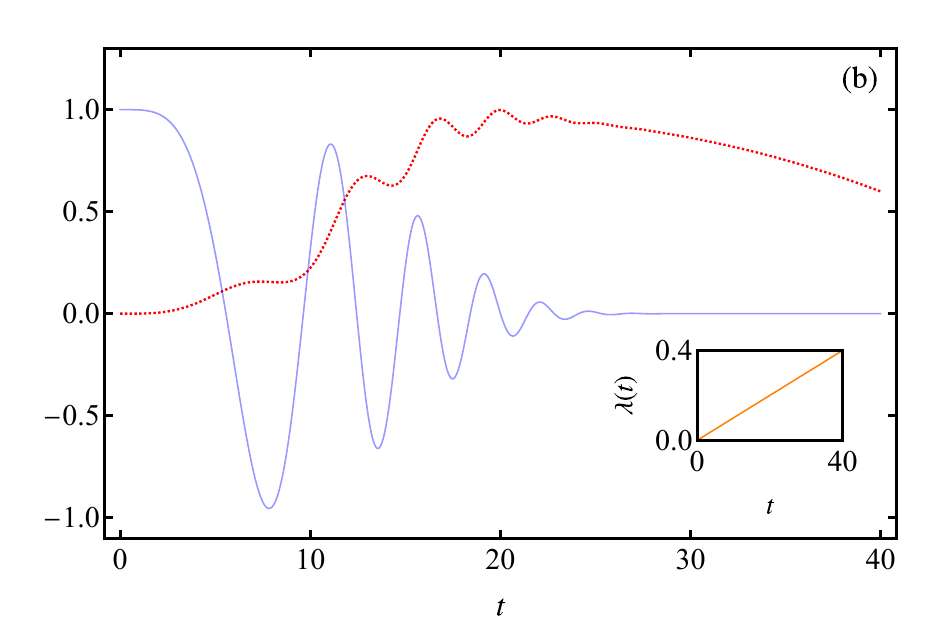}
  \caption{
    The population inversion (solid blue line) and von Neumann entropy
    (dotted red line) as a function of dimensionless time, when
    considering the initial state
    $|\Psi (0) \rangle = |e,\alpha \rangle$, an average photon number of
    $\langle n \rangle = 25$, $\lambda_0=1$ and the linear modulation.
    In (a), we depict the sudden change in the coupling, with
    $\zeta_1=0.16$, while in (b) we represent the adiabatic change,
    $\zeta_1=0.01$.
    As an inset in both plots, we present the behavior of the
    time-dependent coupling.
    We observe a change in the entanglement dynamics, with its minimum
    and maximum occurring at different times compared to the constant
    coupling scenario.
  }
  \label{fig:linear}
\end{figure}

For an initial state  $|\Psi(0)\rangle=|e,\alpha\rangle$,
the population inversion resulting from the linear coupling is
\begin{equation}
    \label{eq:pop_inversion_linear_coupling}
    W(t) = \sum_{n=0}^{\infty} P_n
    \cos \left( \sqrt{n+1}  \lambda_0 \zeta_1 t^2  \right).
\end{equation}
As shown in Fig. \ref{fig:linear}, the quadratic time dependence in the
arguments of the cosine functions in
Eq. \eqref{eq:pop_inversion_linear_coupling} leads to
denser Rabi oscillations than in the standard JC model.
In addition, we observe in Fig. \ref{fig:linear}(a)
(Fig. \ref{fig:linear}(b)) a sudden (adiabatic) change in the coupling,
corresponding to larger (smaller) values of $\zeta_1$, increases
(decreases) the Rabi frequencies, resulting in a denser (less dense)
atomic inversion curve and a shorter (longer) first revival time, now
given by
$t_r=2\sqrt{\pi\langle\hat n\rangle^{1/2}/\zeta_1\lambda _0}$
\cite{Joshi1993}.
Besides, following the shift in the revival time, the behavior of the
von Neumann entropy is also affected.
In the sudden change case, the minimum entanglement is achieved earlier
than in the usual scenario, whereas under adiabatic change, it occurs
later.

\subsection{Hyperbolic Secant}

\begin{figure}[t]
  \centering
  \includegraphics[width=1\linewidth]{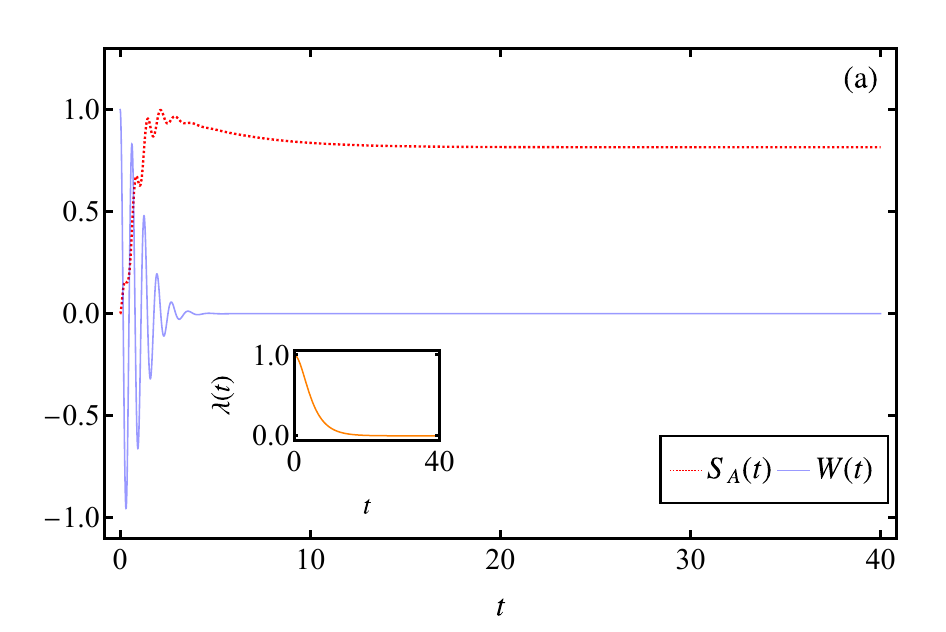}
  \includegraphics[width=1\linewidth]{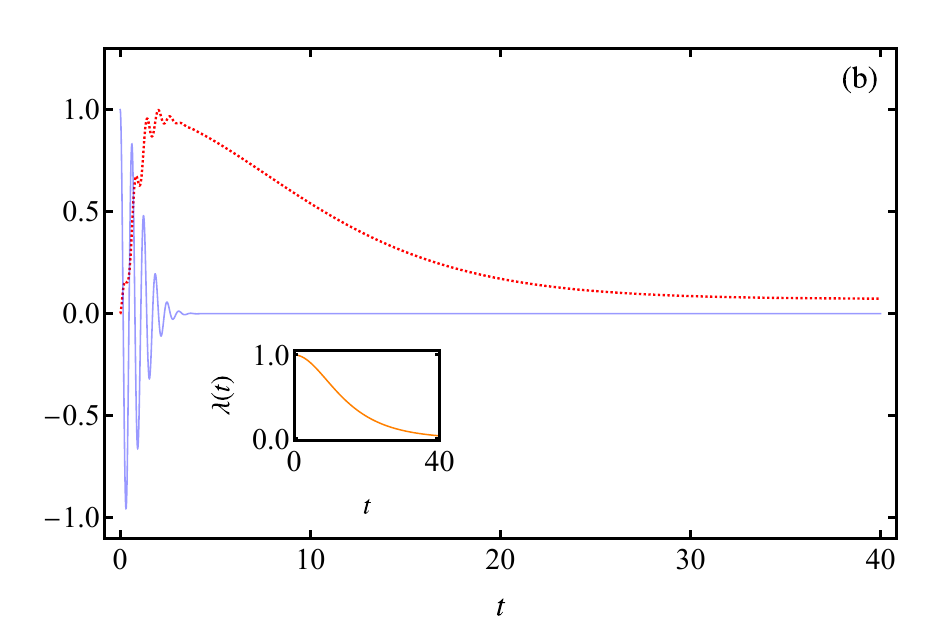}
  \caption{
   The population inversion (solid blue line) and von Neumann entropy
   (dotted red line) as a function of dimensionless time, when
   considering the initial state
   $|\Psi (0) \rangle = |e,\alpha \rangle$, an average photon number of
   $\langle n \rangle = 25$, $\lambda_0=1$ and the hyperbolic secant
   modulation.
    In (a), we present a faster decay in the coupling, with $\zeta_2=0.3$,
    while in (b) we represent a slower stabilization, $\zeta_2=0.1$.
    As an inset in both plots, we present the behavior of the
    time-dependent coupling.
    We observe an asymptotic value for both the population inversion and
    the entanglement.
  }
  \label{fig:sech}
\end{figure}

Dasgupta \cite{Dasgupta1999} suggested the coupling modulation
\begin{equation}
    \lambda(t)=\lambda_0 \sech \left(\zeta_2 t\right),
\end{equation}
This type of coupling can be used to simulate transient effects in the
cavity, allowing the interaction to be switched on and off, depending on
the choice of the initial time.
The population inversion in this scenario is given by
\begin{equation}
    \label{eq:pop_inversion_sech_coupling}
    W(t) = \sum_{n=0}^{\infty} P_n \cos \left\{\frac{2 \lambda_0
           \sqrt{n+1}}{\zeta_{2}}
           \tan ^{-1} \left[\sinh (\zeta_2  t) \right] \ \right\}.
\end{equation}

In Fig. \ref{fig:sech}, we observe how this coupling, which decays to
zero, can lead to different asymptotic values for both population
inversion and entanglement.
In Fig. \ref{fig:sech}(a), we set $\zeta_2= 0.3$ (faster decay of
the coupling), causing the dynamics to stop before reaching the revival
time.
In contrast, in Fig. \ref{fig:sech}(b), we adopt a slower decay for the
coupling.
In both cases, the dynamics cease after the collapse, which causes the
asymptotic value of $W(t)$ to be $0$.
However, for another combination of parameters, it would be possible to
obtain different asymptotic values.
Furthermore, another important aspect is that, even though the
asymptotic value of the population inversion is the same in both cases,
the von Neumann entropy value is significantly different.
The coherence terms of the density matrix, which do not affect the
atomic probabilities are influenced by the different atom-field
decoupling times, and this is reflected in the entanglement behavior.

This concludes the tutorial part of this work.
In the following section, we examine aspects of the atomic dynamics
considering a sinusoidally varying coupling parameter and assuming a
thermal initial state for the cavity field.

\section{Atomic motion}
\label{sec:atomic_mov}

As mentioned in the previous section, Schlicher was the first to
introduce a time-dependent coupling parameter.
In that context, he used a sinusoidal modulation to account for the
motion of the atom along the standing-wave \cite{Schlicher1989}.
Since then, this type of time dependence has been extensively studied:
Bartzis investigated the squeezing in the cavity's degree of freedom
\cite{Bartzis1992}, Wilkens and Meystre extended the approach to maser
systems \cite{Wilkens1992}, and Fang computed the von Neumann entropy
and the Husimi function \cite{Fang1998}, considering an initial coherent
state of the cavity.
Other works onatomic motion in the model can be found in
Refs. \cite{Palma1992,Ren1992,Xie2009,Abdalla2003,Abdel-Khalek2015}.
In this case, we adopt a semiclassical approach, in which atomic
motion is treated classically -- an approximation justified when the
kinetic energy of the atom significantly exceeds the interaction energy
of the JC model \cite{LARSON2021}.

In cavity quantum electrodynamics experiments, the atom crossing the
cavity goes through the spatial profile of the EM field mode, being
subjected to a time-dependent coupling.
To simplify the analysis of the experiments, it is introduced a constant
effective interaction time $t_{eff}$ obtained by averaging the spatial
variation of the atom–field coupling as the atom crosses the cavity
mode.
Thus, the atom is treated as if it traversed the cavity in a time
$t_{eff}$ inversely proportional to the atom's velocity while
experiencing the maximum atom–field coupling
\cite{Haroche1989,Raimond2001,Haroche2006}.
This is justified because, at resonance, the atom–field Hamiltonians at
different times commute, and the time-ordered exponential of the
evolution operator reduces to a simple exponential \cite{Haroche2006}.
The precise control of the atom-field interaction time was essential for
the first experimental observation of collapses and revivals,
accomplished by Haroche’s group \cite{Brune1996}.
For a didactic overview of the experimental methods related to the JC
model in electrodynamic cavities, we recommend Ref. \cite{Haroche2006}.
In the present work, we treat the time-dependent coupling explicitly in
the dynamical evolution of the system, as we are interested in the
effects of such time-varying interaction.

We next address the situation of having the field initially in a thermal
state.
The study of the JC model considering such a state \cite{Glauber1963c}
was first conducted by Cummings \cite{Cummings1965} and later explored
in greater depth by Knight and Radmore \cite{Knight1982}.
This corresponds to a partially cooled cavity \cite{Chumakov1993}, which
is a more realistic scenario given the difficulty of cooling a cavity
very close to $T=0$ K \cite{REMPE1987}.
Unlike the generally well-defined collapses and revivals observed in the
case of an initial coherent state, for a thermal state, the population
inversion exhibits an erratic time evolution due to large photon number
fluctuations.
Later, Arancibia-Bulnes et al., \cite{Arancibia-Bulnes1993} and Azuma
\cite{Azuma2008} investigated the evolution of the Bloch vector in the
JC model with an initial thermal field, the latter concluding that the
``thermal JC model'' could serve as a future source of entanglement.
Additional studies on the model with an initial thermal state, but
without taking into account the atomic motion, can be found in Refs.
\cite{Aravind1984,Foerster1975,Buck1981,Sukumar1981,Chumakov1993,
Arroyo-Correa1990,Klimov1999,Bose2001,Scheel2003}.

The combination of atomic motion with an initial thermal field state has
been explored less frequently:
Yan studied two successive atoms passing a cavity and observed
entanglement sudden death \cite{Yan2009}, while Joshi investigated the
dependence of population inversion on detuning, for different initial
states of the atom, with a focus on how the TLS follows the field
adiabatically \cite{Joshi2004}.

In this work, we first analyze the time evolution of the population
inversion, examining the effects of both the periodicity of the
sinusoidal coupling parameter and the significant uncertainty associated
with the thermal state.
The coupling parameter is modeled as
\begin{equation}
  \label{eq:sinusoidal_coup}
    \lambda(t)=\lambda_0 \sin \left(\frac{p \pi v t}{L} \right),
\end{equation}
where $p$ represents the number of half-wave lengths of the field mode,
$v$ is the atomic velocity and $L$ is the cavity length.
Henceforth, we assume that the velocity is $v=\zeta_3 L/\pi$.
In this context, the coupling area, Eq. \eqref{eq:coupling_area}, is
\begin{equation}
  \label{eq:coup_area_sin}
    A(t)= \frac{\lambda_0  \left[1-\cos(p \zeta_3 t)\right]}{p \zeta_{3} }.
\end{equation}
Thus, having $\lambda t$ replaced by the periodic function $A(t)$ in the
time evolution operator,  we expect the system to exhibit periodic
behavior, as shown in Ref. \cite{Fang1998}.
This may have a significant impact on the time evolution of the atomic
population inversion, for instance, as we are going to show below.
We remark that the parameters $p$ and $\zeta_3$ affect the system's
dynamics in the same way, even though they represent different
quantities.

\begin{figure}[t]
  \centering
  \includegraphics[width=1\linewidth]{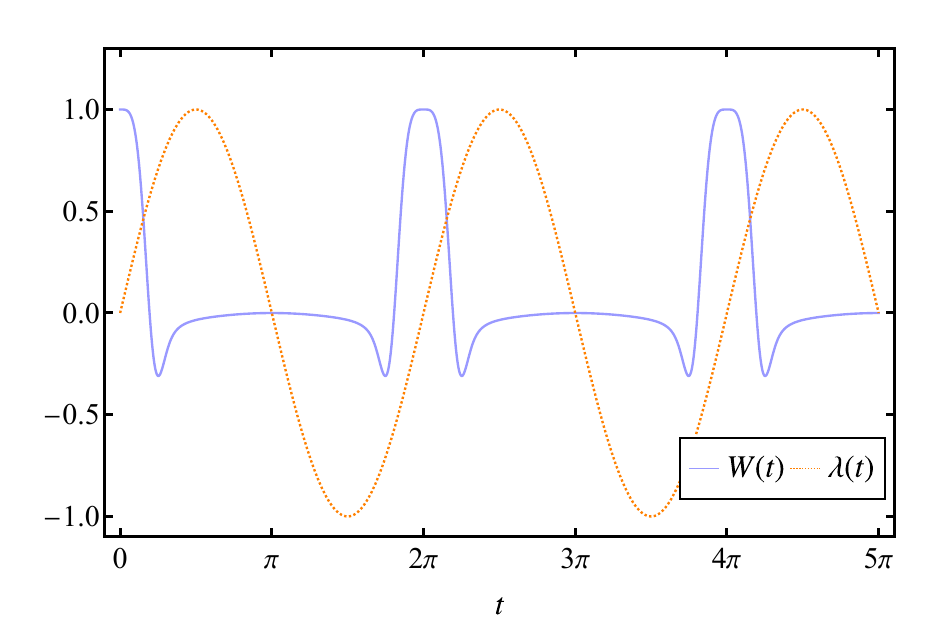}
  \caption{
   The population inversion (solid blue line) and the coupling parameter
   (dotted orange line) as a function of dimensionless time, when
   considering the initial state excited state of the TLS and thermal
   state of the cavity, an average photon number of
   $\langle n \rangle = 25$, $\lambda_0=1$, $\zeta_{3}=1$, $p=1$ and the
   velocity $v=\zeta_{3} L/\pi$.
    As a consequence of the sinusoidal coupling, we observe a periodic
    evolution for the population inversion, even considering a strong
    initial thermal cavity field state.
  }
  \label{fig:thermal}
\end{figure}

Given that the initial field state of the system is mixed, we must
employ the density operator formalism.
We assume the TLS initially in the excited state,
$\hat{\rho}_A(0)=|e\rangle\langle  e|$, and write the initial thermal
field state as \cite{vonNeumann1927b}
\begin{equation}
  \label{eq:thermal_state}
    \hat{\rho}_F(0) = \sum_{n=0}^{\infty} P_n |n\rangle\langle n|,
\end{equation}
where
\begin{equation} \label{eq:thermal_distr}
    P_n=\frac{\expval{n}^n}{\left(1+\expval{n}\right)^{n+1}},
\end{equation}
is the Bose-Einstein distribution.
The average photon number $\expval{n} =[\exp(\nu / k_{B} T)-1]^{-1}$ can
be expressed in terms of the frequency $\nu$ and the effective
temperature $T$, where $k_{B}$ denotes the Boltzmann constant.
The joint atom-field state evolves according to
\begin{equation} \label{eq:evolving_rho}
   \hat{\rho}(t)=\hat{U}(t) \hat{\rho}(0) \hat{U}^\dagger(t),
\end{equation}
where $\hat{U}(t)$ is the time evolution operator given in Eq. \eqref{eq:time_dep_coup_time_ev_op_explicit_all}.
Additionally, we consider an initially separable state of the system  $\hat{\rho}(0)=\hat{\rho}_A(0)\otimes\hat{\rho}_F(0)$.
Following Eqs. \eqref{eq:evolving_rho}, \eqref{eq:Trace} and
\eqref{eq:coup_area_sin}, considering the initial excited atomic state,
we obtain the evolved atomic density matrix
\cite{Boukobza2005,GERRY2005}
\begin{align}
  \label{eq:atomic_z}
  \hat{\rho}_A(t) =  {}
  &
    \sum_{n=0}^\infty P_n \left\{ \cos^2\left[A(t) \sqrt{n+1}\right]
    |e\rangle\langle e| \right. \nonumber \\
  &
    + \left. \sin^2\left[A(t) \sqrt{n+1}\right] |g\rangle\langle g| \right\}.
\end{align}
Therefore, the population inversion in this scenario is explicitly given by
\begin{equation}
    \label{eq:pop_inversion_sinusoidal}
    W(t) = \sum_{n=0}^{\infty} P_n
    \cos \left\{\frac{2 \lambda_0 \sqrt{n+1} }{p \zeta_3}
      \left[ 1 - \cos \left(p \zeta_3 t \right) \right] \right\},
\end{equation}
where we have employed Eqs. \eqref{eq:sinusoidal_coup} and
\eqref{eq:coupling_area}, and $P_n$ corresponds to the Bose-Einstein
distribution, Eq. \eqref{eq:thermal_distr}.
Recall that the parameter $\zeta_3$ is related to the atomic velocity
$v$ through $v=\zeta_3 L/\pi$.
The time evolution of the population inversion is presented in
Fig. \ref{fig:thermal}.
We observe that, rather than the erratic behavior associated with the
thermal field state, the system exhibits periodicity, with a period that
depends on the parameters $p$ and $\zeta_3$.
Additionally, secondary oscillations arise due to inflections in the
coupling parameter, and if we compare the results with
Ref. \cite{Fang1998} for an initial coherent state, we note that the
thermal state gives rise to RO with smaller amplitude, as expected when
considering the constant coupling scenario \cite{Knight1982}.

\begin{figure}[t]
  \centering
  \includegraphics[width=1\linewidth]{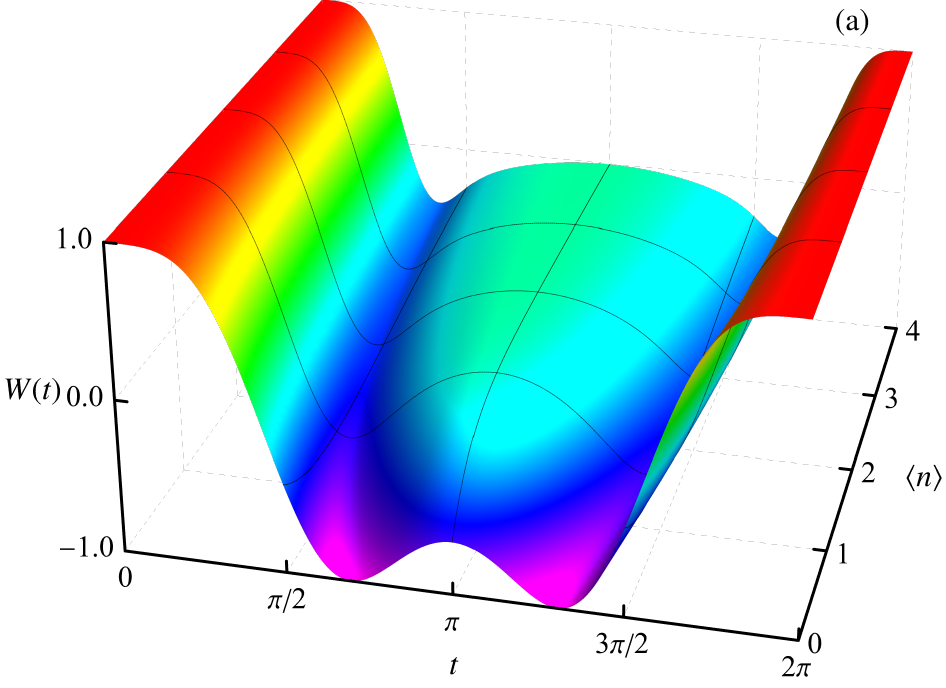}
  \includegraphics[width=1\linewidth]{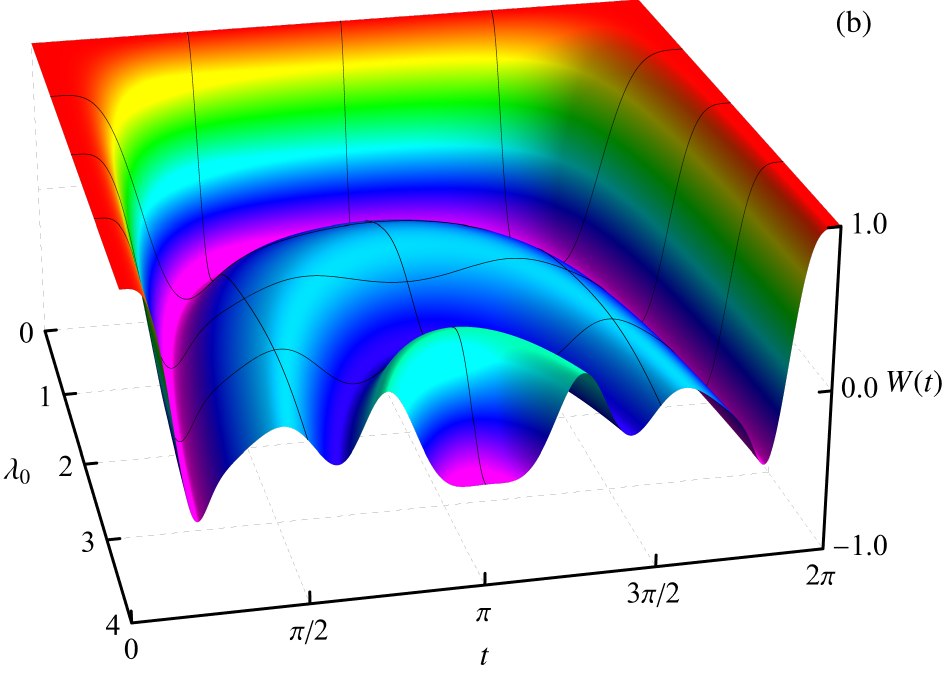}
  \caption{
    In (a), we present a surface plot of the population inversion as a
    function of dimensionless time and the average number of photons,
    $\langle n \rangle$, with $\lambda_0 = 1$ fixed. An increase in
    $\langle n \rangle$ reduces the amplitude while enlarging the small
    concavity at the bottom.
    In (b), we present a surface plot of the population inversion as a
    function of dimensionless time and the parameter $\lambda_0$, with
    $\langle n \rangle = 4$ fixed. An increase in $\lambda_0$ alters the
    oscillation shape.
    In both plots, we set $p = \zeta_3= 1$, and $v=\zeta_{3} L/\pi$.
  }
  \label{fig:surfaceplot}
\end{figure}

In Fig. \ref{fig:surfaceplot} we illustrate how the population inversion
depends on relevant parameters of the system.
In Fig. \ref{fig:surfaceplot}(a) we have plotted $W(t)$ as a function of
the initial average photon number, $\langle n \rangle$, and the
dimensionless time.
Although the period does not depend on $\langle n \rangle$, the
concavity between adjacent two main peaks becomes wider as $\langle n
\rangle$ increases, while its amplitude decreases.
This occurs because a smaller $\langle n \rangle$ sharpens the
probability distribution, which is reflected in the atomic inversion.
Similarly, as shown in Fig. \ref{fig:surfaceplot}(b), while the period
of the main oscillations is independent of the amplitude $\lambda_0$, we
clearly see the emergence of the secondary oscillations due to the
inflections in $\lambda(t)$, which become more pronounced as $\lambda_0$
increases.

\subsection{Bloch vector}
If the global state of the system is mixed, the von Neumann entropy is
no longer a suitable measure of entanglement
\cite{GERRY2005,Nielsen2010}.
Considerable attention has been given to entanglement measures in this
context \cite{Scheel2003,Bose2001,Boukobza2005}, such as the concurrence
in a system of two moving two-level atoms \cite{Yan2009}.
Alternatively, we focus here on the Bloch vector
\cite{Gea-Banacloche1992}, which contains comprehensive information
about the quantum state, including its purity and the degree of
atom–field alignment.
The Bloch vector $\boldsymbol{R}(t)$ is a convenient geometrical
representation of the quantum states of two-level systems.
It is such that $|\boldsymbol{R}(t)| \leq 1$ and can be associated with
points in a unit sphere, known as \emph{Bloch sphere}.
While pure quantum states lie on the sphere's surface,
$|\boldsymbol{R}(t)| = 1$, for mixed states $|\boldsymbol{R}(t)| <
1$.
As any $2 \times 2$ matrix can be expressed as a linear combination of
the Pauli matrices and the identity matrix $\boldsymbol{\hat{1}}_2$,
the reduced atomic density matrix can be parametrized as
\begin{equation}
  \hat{\rho}_A(t)=\frac{1}{2}[\mathbf{\hat{1}}_2
  +\boldsymbol{\hat{\sigma}}\cdot  \mathbf{R}(t)].
\end{equation}
Expressing the Bloch vector in terms of its components as
\begin{equation}
    \mathbf{R}(t)=\left[R_x (t),R_y(t),R_z(t)\right],
\end{equation}
we obtain
\begin{align}
  \label{eq:bloch_components}
  R_x (t) = {} & \langle \hat{\sigma}_x\rangle= \rho_A^{eg}(t)+\rho_A^{eg}(t),\nonumber\\
  R_y (t) = {} &  i\left[\rho_A^{eg}(t)-\rho_A^{ge}(t)\right],\nonumber\\
  R_z (t) = {} & W(t) = \rho _A^{ee}(t)-\rho _A^{gg}(t),
\end{align}
where $\rho _A^{ij}$ represents the elements of $\hat{\rho}_A(t)$, with
$i=e,g$.
Interpretation-wise, the component $R_x(t)$ corresponds to the
expectation value of the Pauli matrix in the $x$-direction, also known
as the dipole operator, which quantifies the atom's alignment with the
field \cite{LARSON2021}.
It can be expressed by
$\hat{\sigma}_x=\hat{\sigma}_++\hat{\sigma}_-$.
Moreover, $R_z(t)$ represents the population inversion, while
$R(t) = |\mathbf{R}(t)|$ serves as a measure of atomic purity
\cite{Gea-Banacloche1992}.
It follows that $R(t)$ is inversely related to the von Neumann entropy,
as the latter quantifies the mixedness of the atomic subsystem.
We choose to focus on the Bloch vector -- its motion, components, and
modulus -- for comparison with the results mentioned in the next
paragraph.

\begin{figure*}
  \centering
  \includegraphics[width=0.9\linewidth]{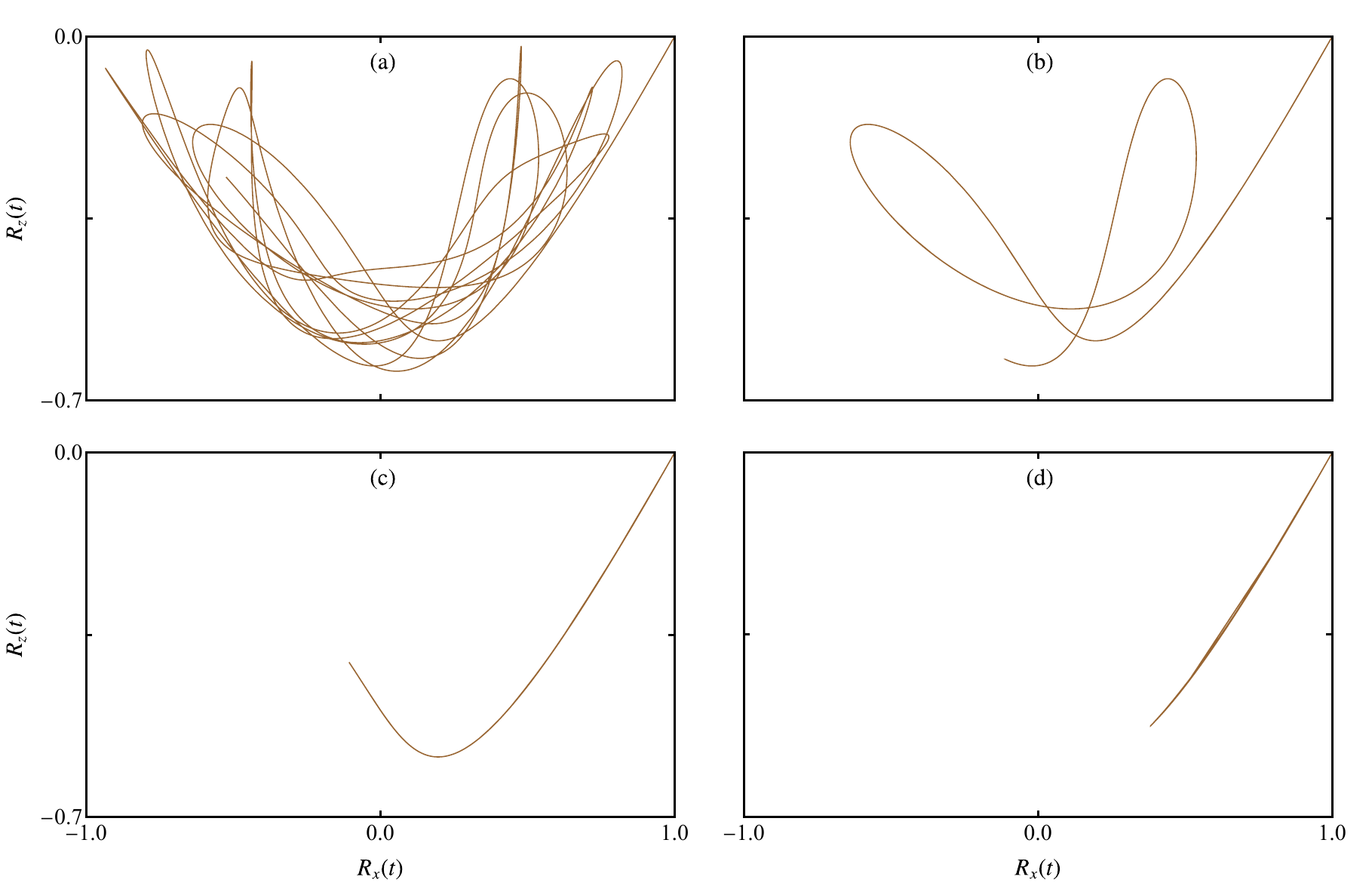}
  \caption{
    Parameter plot of the
    $R_x(t)$ and $R_z(t)$ components as a function of time.
    In panel (a), atomic motion is not considered, and we set
    the parameter $\lambda = 1$.
    Conversely, in the remaining panels, atomic motion is included, with
    parameters $\lambda_0 = 1$, $v = \zeta_3 \pi / L$ and considered a
    time interval $0 < t < 40$.
    In panels (b) and (c), we set $p = 1$, whereas in panel
    (d), we use $p = 2$.
    In (b), we employ $\zeta_3=0.25$, while in the other panels,
    we fix $\zeta_3=1$.
    In all cases, the initial average photon number is
    $\langle n \rangle = 0.5$.
    We observe that atomic motion dominates the disordered behavior of
    the Bloch vector.
  }
      \label{fig:bloch1}
\end{figure*}

\begin{figure*}[t]
  \centering
  \includegraphics[width=0.9\linewidth]{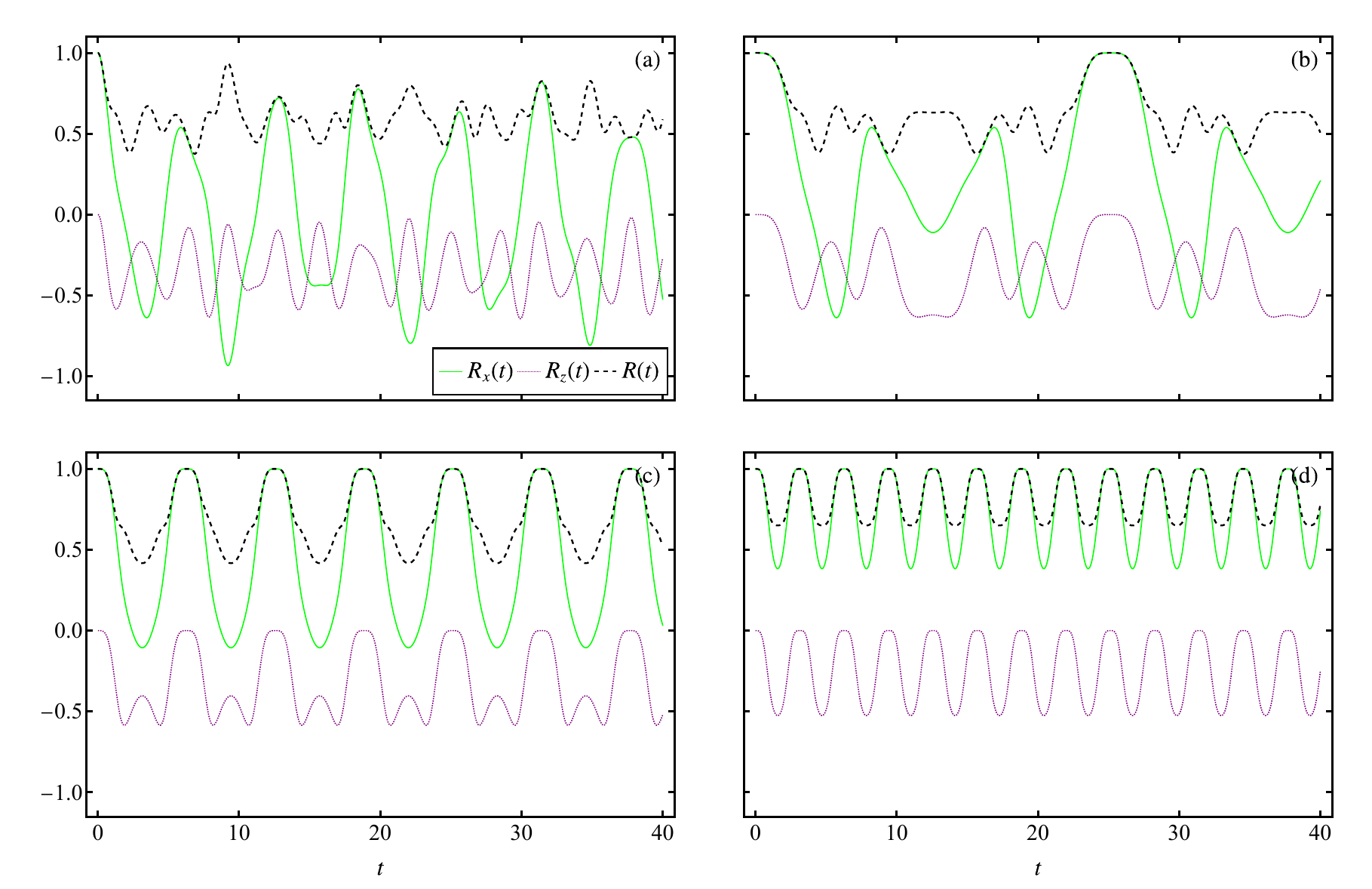}
  \caption{
    Time-evolution of the components of the Bloch vector: $R_x(t)$
    (solid green line), $R_z(t)$ (dotted purple line), and $R(t)$
    (dashed black line).
    In panel (a), atomic motion is not considered, and we set
    the parameter $\lambda = 1$.
    Conversely, in the remaining panels, atomic motion is included, with
    parameters $\lambda_0 = 1$, $v = \zeta_3 \pi / L$ and considered a
    time interval $0 < t < 40$.
    In panels (b) and (c), we set $p = 1$, whereas in panel
    (d), we use $p = 2$.
    In (b), we employ $\zeta_3=0.25$, while in the other panels,
    we fix $\zeta_3=1$.
    In all cases, the initial average photon number is
    $\langle n \rangle = 0.5$.
    Additionally, an increase in $p$ diminishes the amplitude of
    $R_x(t)$ and $R(t)$.
  }
      \label{fig:bloch2}
\end{figure*}

The dynamics of the Bloch vector in the context of the JC model was
first studied by Gea-Banacloche in 1992 \cite{Gea-Banacloche1992},
considering an initial coherent state in the cavity.
As mentioned earlier, this analysis was later extended considering to
the case of an initial thermal state, revealing erratic Bloch vector
dynamics \cite{Arancibia-Bulnes1993,Azuma2008}.
In a subsequent study, Azuma and Ban demonstrated that this behavior
exhibits a quasiperiodic structure \cite{Azuma2014}.
Joshi and Xiao studied the simultaneous effects of atomic motion with an
initial thermal state, considering detuning and the adiabatic
approximation, and focusing on transparency effects \cite{Joshi2004}.
Here, we explore the influence of atomic motion on the evolution of the
Bloch vector, considering the cavity initially in a thermal state.

To analyze the dynamics of the Bloch vector in the $xz$ plane, we modify
the initial TLS state to an eigenvector of $\hat{\sigma}_x$, namely
$|\psi_A(0)\rangle=1/\sqrt{2}(|e\rangle+|g\rangle)$.
In the density operator formalism, the initial state of the system is,
then,
\begin{equation}
  \label{eq:initial_dipole}
  \hat{\rho}(0)=
  \left[\frac{1}{2}\left(|e\rangle\langle e|
      +|e\rangle\langle g|
      +|g\rangle\langle g|
      +|g\rangle\langle e|\right)\right]\otimes \hat{\rho}^F(0),
\end{equation}
where $\hat{\rho}_F(0)$ is described by Eq. \eqref{eq:thermal_state}.
This contrasts with the usual choice for the initial state,
$\hat{\rho}_A(0)=|e\rangle\langle e|$, which yields a diagonal atomic
density matrix without coherence terms, restricting the Bloch vector to
the $z$-axis, as seen in Eq. \eqref{eq:atomic_z}.
Additionally, this change facilitates comparison with previous results.
Employing Eqs. \eqref{eq:Trace},
\eqref{eq:time_dep_coup_time_ev_op_explicit_all} and
\eqref{eq:evolving_rho},  and considering the initial state in Eq.
\eqref{eq:initial_dipole}, we obtain the evolved atomic density matrix
\begin{align}
  \hat{\rho}_{A}(t) = {}
  &
    \frac{1}{2} \sum_{n=0}^{\infty} P_n
    \left\{ \cos^{2}\left[A(t) \sqrt{n+1}\right]
    + \sin^{2}\left[A(t) \sqrt{n}\right] \right\} |e\rangle \langle e|
    \nonumber \\
  &
    + \frac{1}{2} \sum_{n=0}^{\infty} P_n
    \cos\left[A(t) \sqrt{n+1}\right]
    \cos\left[A(t) \sqrt{n}\right] |e\rangle \langle g|
    \nonumber \\
  &
    + \frac{1}{2} \sum_{n=0}^{\infty} P_n
    \cos\left[A(t) \sqrt{n}\right]
    \cos\left[A(t) \sqrt{n+1}\right] |g\rangle \langle e|
    \nonumber \\
  &
    +  \frac{1}{2} \sum_{n=0}^{\infty} P_n
    \left\{ \cos^{2}\left[A(t) \sqrt{n}\right]
    + \sin^{2}\left[A(t) \sqrt{n+1}\right] \right\} |g\rangle \langle g|.
\end{align}
From this result, we can calculate the components of the Bloch vector,
according to Eq. \eqref{eq:bloch_components},
\begin{align}
  R_{x}(t) = {}
  &
    \sum_{n=0}^{\infty} P_{n} \cos\left[A(t) \sqrt{n}\right]
    \cos\left[A(t) \sqrt{n+1}\right], \nonumber \\
  R_y(t) =  {}
  &
    0, \nonumber \\
  R_z(t) =  {}
  & \sum_{n=0}^{\infty} P_{n} \left\{
    \cos^{2}\left[A(t)\sqrt{n+1}\right] +
    \sin^{2}\left[A(t)\sqrt{n}\right] \right\}-1.
\end{align}

In Fig. \ref{fig:bloch1}, we plot the components $R_x(t)$ and $R_z(t)$
as functions of time in a parameter plot.
In the constant coupling case, shown in panel (a) of
Fig. \ref{fig:bloch1}, we set $\lambda = 1$.
For the sinusoidal coupling case, depicted in the other panels of
Fig. \ref{fig:bloch1}, we fix the parameters as $\lambda_0 = 1$ and
$v=\zeta_3 L/\pi$.
In panels (b) and (c) we set $p = 1$, while in panel (d), we set $p =
2$.
In (b), we employ $\zeta_3=0.25$, while in the other panels, we
fix $\zeta_{3}=1$.
At the outset, we choose a low average photon number,
$\langle n \rangle = 0.5$, corresponding to a low temperature, for all
scenarios.
Comparing panels (a) with (b), (c) and (d) of Fig. \ref{fig:bloch1}, we
observe that the sinusoidal  coupling introduces periodicity into the
Bloch vector's motion, replacing the disordered behavior.
In contrast, for a smaller $\zeta_3$, as shown in (b), the behavior
partially resembles that of the time-independent coupling case,
exhibiting reduced-amplitude oscillations within each period.
This suggests that for lower values of the parameter $\zeta_3$, an
intermediate regime emerges in which vestiges of the aperiodic evolution
persist.

In Fig. \ref{fig:bloch2}, we present the time-evolution of the
components and absolute value of the Bloch vector for both constant and
sinusoidal coupling scenarios.
The parameters in Fig. \ref{fig:bloch2}(a)-(d) correspond to those in
Fig. \ref{fig:bloch1}(a)-(d), respectively.
In Fig. \ref{fig:bloch2}(b), (c) and (d), we again observe the
periodicity inherited from the modulated coupling, as compared to
Fig. \ref{fig:bloch2}(a).
Furthermore, as shown in panels (b) and (c), the increase to $\zeta_3=1$
results in a higher frequency of the oscillations.
Additionally, the amplitudes of the oscillations in $R_x(t)$ decrease,
whereas $R_z(t)$
oscillates with approximately the same amplitude.

Furthermore, a comparison of Figs. \ref{fig:bloch2}(c) and (d) shows
that an increase in the number of half wavelengths, $p$, not only leads
to faster oscillations, but also reduces the amplitude of $R_x(t)$ and
$R(t)$, while notably increasing the time-averaged values of these
quantities. This is because a larger $p$ favors the alignment of the
atom with the field, thus increasing the mean value of $R_x(t)$ and also
the mean purity of the atomic state, here quantified by $R(t)$.
We also note a subtle decrease in the amplitude of the oscillations in
$R_z(t)$, indicating a weaker influence of the cavity on the atomic
populations.
This effect can also be attributed to the enhanced alignment of the atom
with the field.

In all examples with atomic motion, we begin with $R(0)=R_x(0)=1$ and
$R_z(0)=0$, meaning that initially, the atomic subsystem is pure, the
atomic transition is aligned with the field, and the excited and ground
populations are equal.
The evolution of the system leads to a periodic return to the initial
state.
This behavior is a direct consequence of the trigonometric dependence of
the coupling area $A(t)$, as discussed previously, which can also be
interpreted in terms of Bloch vector rotations for the chosen velocity,
as predicted by Joshi and Xiao \cite{Joshi2004}.

\begin{figure}[b]
  \centering
  \includegraphics[width=1\linewidth]{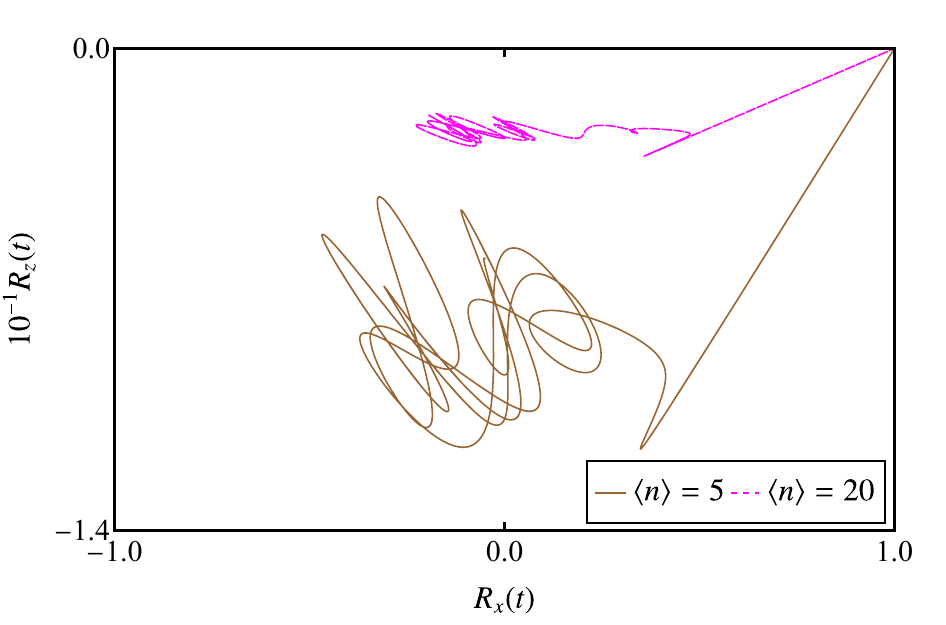}
  \caption{
    Bloch vector trajectories for larger initial average thermal photon
    numbers considering atomic motion with (i) $\langle n \rangle=5$
    (solid brown line) and (ii) $\langle n \rangle=20$ (dashed magenta
    line).
    We set $\lambda_0 = 1$, $p=1$ and $\zeta_3=0.1$ and considered a
    time interval $0 < t < 100$.
  }
  \label{fig:comparison}
\end{figure}

A question that arises is how increasing the mean photon number of the
initial thermal state  affects the Bloch vector dynamics.
On the one hand, sinusoidal modulation of the coupling induces a
periodic evolution of the Bloch vector; on the other, stronger thermal
fluctuations are expected to produce a more disordered evolution.
For instance, considering a larger initial average photon number,  we
expect a reduction in the time-averaged purity due to the increase of
thermal fluctuations.
Nevertheless, increasing $\langle n \rangle$ gives rise to another
effect, a kind of \emph{atomic population trapping}. This is clearly
shown in Fig. \ref{fig:comparison}, where we plot the Bloch vector
trajectory for two different initial mean photon numbers, $\langle n
\rangle = 5$ and $\langle n \rangle = 25$.
Thus, increasing the thermal noise does not significantly enhance the
disorder in the evolution, which remains dominated by the periodic
forcing, but instead tends to freeze the atomic population.
A similar effect occurs in the constant-coupling case, as discussed in
Ref. \cite{Arancibia-Bulnes1993}.
We remark that this kind trapping differs in nature from the population
trapping found in the JC model with an initial coherent state
\cite{Zaheer89,Jonathan1999}.
In the present case, the trapping arises from the cancellation of photon
emission (absorption) by the upper (lower) atomic state
\cite{Arancibia-Bulnes1993}, whereas in the coherent state case it is a
phase-dependent phenomenon \cite{Zaheer89,Jonathan1999}.

Several other modulations can be found in the literature, which the
reader may eventually find useful.
For instance, an alternative to simulate transient effects in the
cavity, the exponential dependence has been considered by Prants
\cite{Prants1992}.
In Ref. \cite{Larson2003}, various distinct coupling pulses, such as
square, Gaussian, and Lorentzian waves, were used to analyze the
perspective of photon filters.
Besides, the sinusoidal time dependence has its generalizations,
including cases of accelerating atoms \cite{Abdalla2003}, where the
argument of the trigonometric function is quadratic in time, or the
alignment of the atom’s dipole moment \cite{Abdel-Khalek2015}, where the
sine function appears squared.
An explicit approach to controlling the population inversion via the
time-dependent coupling parameter is detailed in Ref. \cite{Yang2006}.

\section{Conclusions}
\label{sec:conc}

In the context of the 100-year anniversary of quantum mechanics, the
Jaynes-Cummings model stands as a cornerstone for understanding
light–matter interaction.
More than half a century after its introduction, its exact solvability
and conceptual clarity continue to inspire both theoretical developments
and experimental implementations at the frontiers of quantum science and
technology.
Building on this legacy, we have investigated the resonant
time-dependent Jaynes-Cummings model, where the modulation of the
atom-field coupling parameter may simulate physical phenomena such as
transient behavior and atomic motion.
We presented a pedagogical derivation of a solution of the
time-dependent Jaynes-Cummings model valid for a wide range of initial
conditions, assuming a time-dependent atom-field coupling.
The first type of modulation considered was a linear ramp, associated
with variations in the cavity mode.
We observed that the behavior of entropy and population inversion was
significantly affected by the control parameter $\zeta_1$, the slope of
the ramp, which can delay or advance the evolution of the Rabi
oscillations.
The second type of coupling modulation, given by a hyperbolic secant
function, accounted for transient effects in the cavity, such as
switching on and off.
Our results indicated that, by simulating the gradual turning off of the
cavity, asymptotic values of the atom-field entanglement can be reached,
depending on the control parameter $\zeta_2$, which sets the time scale
of the atom-field coupling decay.
Entanglement plays a central role in quantum information processing, and
these modulations offer a way of controlling it.
Next, we examined another physically relevant scenario in which an atom
moves inside a cavity that is not completely cooled, analyzing the
resulting behavior of the population inversion.
As a consequence of the trigonometric coupling, the Rabi oscillations
became periodic, in contrast to the constant-coupling scenario.
We also investigated how the parameters $\langle n \rangle$ and
$\lambda_0$ affect the population inversion.
The average initial photon number $\langle n \rangle$ influences the
amplitude of the oscillations, also widening the concave region between
neighboring main peaks, while $\lambda_0$ introduces secondary
oscillations due to the inflection in the coupling parameter.

Finally, we analyzed the influence of atomic motion on the dynamics of
the Bloch vector, a quantity that provides an accurate characterization
of the system, considering the atom initially prepared in a
superposition of ground and excited states (an eigenvector of
$\hat\sigma_x$).
Our analysis revealed a number of interesting aspects of the
system. Unlike the disordered dynamics in the constant-coupling
scenario, the periodically varying atom-field coupling parameter leads
to a periodic evolution of the Bloch vector trajectories, dominating
over thermal fluctuations.
However, for sufficiently small values of the control parameter
$\zeta_3$, proportional to the frequency of the coupling modulation, we
observe longer-period oscillations, resembling the dynamics of the
constant-coupling case.
On the other hand, increasing either the number of half-wavelengths in
the field structure, $p$, or the parameter $\zeta_3$, which is
proportional to the atomic velocity, results in a larger time-averaged
purity of the atomic subsystem.
Our results also reveal an interesting underlying competition in the
dynamics: increasing the average thermal photon number, instead of
introducing further disorder into the Bloch vector trajectories, induces
a type of population trapping effect.
As quantum mechanics reaches its centenary, the Jaynes–Cummings model,
over 60 years old yet distinguished by its blend of simplicity and
depth, continues to thrive, advancing our understanding of light–matter
interaction as well as offering useful guidance for applications in
quantum technologies.

\medskip

\section*{Acknowledgments}
We thank Dr. Alison A. Silva for helpful discussions.
This work was financially supported by the Coordenação de Aperfeiçoamento
de Pessoal de Nível Superior (CAPES, Finance Code 001), Conselho
Nacional de Desenvolvimento Científico e Tecnológico (CNPq), Instituto
Nacional de Ciência e Tecnologia de Informação Quântica (INCT-IQ),
Brazil, and the  Air Force Office of Scientific Research (AFOSR), USA.
D.C. would like to acknowledge financial support from Instituto
Serrapilheira, and the Pró-Reitoria de Pesquisa e Inovação (PRPI) from
the Universidade de São Paulo (USP) by financial support through the
Programa de Estímulo à Supervisão de Pós-Doutorandos por Jovens
Pesquisadores.
F.M.A. acknowledges the CNPq Grant No 313124/2023-0, F.M.A. and
A.S.M.C. acknowledge Fundação Araucária Project No 305, and
A.V.-B. acknowledges the AFOSR award No FA9550-24-1-0009.

\medskip

\textbf{Funding}
This work was supported by the Coordenação de Aperfeiçoamento de Pessoal de Nível Superior (CAPES, Finance Code 001), the Conselho Nacional de Desenvolvimento Científico e Tecnológico (CNPq), the Instituto Nacional de Ciência e Tecnologia de Informação Quântica (INCT-IQ), Brazil, and the Air Force Office of Scientific Research (AFOSR), USA.
D.C. was supported by Instituto Serrapilheira and by the Pró-Reitoria de Pesquisa e Inovação (PRPI) from the Universidade de São Paulo (USP) through the Programa de Estímulo à Supervisão de Pós-Doutorandos por Jovens Pesquisadores.
F.M.A. acknowledges support from CNPq (Grant No. 313124/2023-0) and
Fundação Araucária (Project No. 305).
A.S.M.C. acknowledges support from  Fundação Araucária (Project No. 305)
A.V.-B. acknowledges support from AFOSR (Award No. FA9550-24-1-0009).

\medskip

\textbf{Author Contributions}
The work is partly the result of T.T.T.’s PhD thesis under the
supervision of F.M.A. and A.S.M.C..
The authors performed all calculations, figures, and graphs in
partnership.
T.T.T. and D.C. calculated the quantities in the dynamics.
T.T.T. and A.V.-B. wrote the main manuscript text.
All authors reviewed the manuscript.

\medskip

\textbf{Conflict of Interest}
The authors declare no competing interests.

\medskip

\textbf{Data Availability}
No datasets were generated or analyzed during the current study.

\bibliographystyle{apsrev4-2}

\begin{thebibliography}{114}%
\makeatletter
\providecommand \@ifxundefined [1]{%
 \@ifx{#1\undefined}
}%
\providecommand \@ifnum [1]{%
 \ifnum #1\expandafter \@firstoftwo
 \else \expandafter \@secondoftwo
 \fi
}%
\providecommand \@ifx [1]{%
 \ifx #1\expandafter \@firstoftwo
 \else \expandafter \@secondoftwo
 \fi
}%
\providecommand \natexlab [1]{#1}%
\providecommand \enquote  [1]{``#1''}%
\providecommand \bibnamefont  [1]{#1}%
\providecommand \bibfnamefont [1]{#1}%
\providecommand \citenamefont [1]{#1}%
\providecommand \href@noop [0]{\@secondoftwo}%
\providecommand \href [0]{\begingroup \@sanitize@url \@href}%
\providecommand \@href[1]{\@@startlink{#1}\@@href}%
\providecommand \@@href[1]{\endgroup#1\@@endlink}%
\providecommand \@sanitize@url [0]{\catcode `\\12\catcode `\$12\catcode
  `\&12\catcode `\#12\catcode `\^12\catcode `\_12\catcode `\%12\relax}%
\providecommand \@@startlink[1]{}%
\providecommand \@@endlink[0]{}%
\providecommand \url  [0]{\begingroup\@sanitize@url \@url }%
\providecommand \@url [1]{\endgroup\@href {#1}{\urlprefix }}%
\providecommand \urlprefix  [0]{URL }%
\providecommand \Eprint [0]{\href }%
\providecommand \doibase [0]{https://doi.org/}%
\providecommand \selectlanguage [0]{\@gobble}%
\providecommand \bibinfo  [0]{\@secondoftwo}%
\providecommand \bibfield  [0]{\@secondoftwo}%
\providecommand \translation [1]{[#1]}%
\providecommand \BibitemOpen [0]{}%
\providecommand \bibitemStop [0]{}%
\providecommand \bibitemNoStop [0]{.\EOS\space}%
\providecommand \EOS [0]{\spacefactor3000\relax}%
\providecommand \BibitemShut  [1]{\csname bibitem#1\endcsname}%
\let\auto@bib@innerbib\@empty
\bibitem [{\citenamefont {Jaynes}\ and\ \citenamefont
  {Cummings}(1963)}]{JAYNES1963}%
  \BibitemOpen
  \bibfield  {author} {\bibinfo {author} {\bibfnamefont {E.~T.}\ \bibnamefont
  {Jaynes}}\ and\ \bibinfo {author} {\bibfnamefont {F.~W.}\ \bibnamefont
  {Cummings}},\ }\href {https://doi.org/10.1109/PROC.1963.1664} {\bibfield
  {journal} {\bibinfo  {journal} {Proc. IEEE}\ }\textbf {\bibinfo {volume}
  {51}},\ \bibinfo {pages} {89} (\bibinfo {year} {1963})}\BibitemShut {NoStop}%
\bibitem [{\citenamefont {Larson}\ \emph {et~al.}(2024)\citenamefont {Larson},
  \citenamefont {Mavrogordatos}, \citenamefont {Parkins},\ and\ \citenamefont
  {Vidiella-Barranco}}]{Larson2024}%
  \BibitemOpen
  \bibfield  {author} {\bibinfo {author} {\bibfnamefont {J.}~\bibnamefont
  {Larson}}, \bibinfo {author} {\bibfnamefont {T.}~\bibnamefont
  {Mavrogordatos}}, \bibinfo {author} {\bibfnamefont {S.}~\bibnamefont
  {Parkins}},\ and\ \bibinfo {author} {\bibfnamefont {A.}~\bibnamefont
  {Vidiella-Barranco}},\ }\href {https://doi.org/10.1364/JOSAB.536847}
  {\bibfield  {journal} {\bibinfo  {journal} {J. Opt. Soc. Am. B}\ }\textbf
  {\bibinfo {volume} {41}},\ \bibinfo {pages} {JCM1} (\bibinfo {year}
  {2024})}\BibitemShut {NoStop}%
\bibitem [{\citenamefont {{De Bernardis}}\ \emph {et~al.}(2024)\citenamefont
  {{De Bernardis}}, \citenamefont {Mercurio},\ and\ \citenamefont {{De
  Liberato}}}]{DeBernardis2024}%
  \BibitemOpen
  \bibfield  {author} {\bibinfo {author} {\bibfnamefont {D.}~\bibnamefont {{De
  Bernardis}}}, \bibinfo {author} {\bibfnamefont {A.}~\bibnamefont
  {Mercurio}},\ and\ \bibinfo {author} {\bibfnamefont {S.}~\bibnamefont {{De
  Liberato}}},\ }\href {https://doi.org/10.1364/JOSAB.522786} {\bibfield
  {journal} {\bibinfo  {journal} {J. Opt. Soc. Am. B}\ }\textbf {\bibinfo
  {volume} {41}},\ \bibinfo {pages} {C206} (\bibinfo {year}
  {2024})}\BibitemShut {NoStop}%
\bibitem [{\citenamefont {Cummings}(2013)}]{Cummings2013}%
  \BibitemOpen
  \bibfield  {author} {\bibinfo {author} {\bibfnamefont {F.~W.}\ \bibnamefont
  {Cummings}},\ }\href {https://doi.org/10.1088/0953-4075/46/22/220202}
  {\bibfield  {journal} {\bibinfo  {journal} {J. Phys. B}\ }\textbf {\bibinfo
  {volume} {46}},\ \bibinfo {pages} {220202} (\bibinfo {year}
  {2013})}\BibitemShut {NoStop}%
\bibitem [{\citenamefont {Cummings}(1965)}]{Cummings1965}%
  \BibitemOpen
  \bibfield  {author} {\bibinfo {author} {\bibfnamefont {F.~W.}\ \bibnamefont
  {Cummings}},\ }\href {https://doi.org/10.1103/PhysRev.140.A1051} {\bibfield
  {journal} {\bibinfo  {journal} {Phys. Rev.}\ }\textbf {\bibinfo {volume}
  {140}},\ \bibinfo {pages} {A1051} (\bibinfo {year} {1965})}\BibitemShut
  {NoStop}%
\bibitem [{\citenamefont {Meystre}\ \emph {et~al.}(1975)\citenamefont
  {Meystre}, \citenamefont {Geneux}, \citenamefont {Quattropani},\ and\
  \citenamefont {Faist}}]{Meystre1975}%
  \BibitemOpen
  \bibfield  {author} {\bibinfo {author} {\bibfnamefont {P.}~\bibnamefont
  {Meystre}}, \bibinfo {author} {\bibfnamefont {E.}~\bibnamefont {Geneux}},
  \bibinfo {author} {\bibfnamefont {A.}~\bibnamefont {Quattropani}},\ and\
  \bibinfo {author} {\bibfnamefont {A.}~\bibnamefont {Faist}},\ }\href
  {https://doi.org/10.1007/BF02724735} {\bibfield  {journal} {\bibinfo
  {journal} {Il Nuovo Cimento B}\ }\textbf {\bibinfo {volume} {25}},\ \bibinfo
  {pages} {521} (\bibinfo {year} {1975})}\BibitemShut {NoStop}%
\bibitem [{\citenamefont {Eberly}\ \emph {et~al.}(1980)\citenamefont {Eberly},
  \citenamefont {Narozhny},\ and\ \citenamefont
  {Sanchez-Mondragon}}]{Eberly1980}%
  \BibitemOpen
  \bibfield  {author} {\bibinfo {author} {\bibfnamefont {J.~H.}\ \bibnamefont
  {Eberly}}, \bibinfo {author} {\bibfnamefont {N.~B.}\ \bibnamefont
  {Narozhny}},\ and\ \bibinfo {author} {\bibfnamefont {J.~J.}\ \bibnamefont
  {Sanchez-Mondragon}},\ }\href {https://doi.org/10.1103/PhysRevLett.44.1323}
  {\bibfield  {journal} {\bibinfo  {journal} {Phys. Rev. Lett.}\ }\textbf
  {\bibinfo {volume} {44}},\ \bibinfo {pages} {1323} (\bibinfo {year}
  {1980})}\BibitemShut {NoStop}%
\bibitem [{\citenamefont {Meschede}\ \emph {et~al.}(1985)\citenamefont
  {Meschede}, \citenamefont {Walther},\ and\ \citenamefont
  {M{\"{u}}ller}}]{MESCHEDE1985}%
  \BibitemOpen
  \bibfield  {author} {\bibinfo {author} {\bibfnamefont {D.}~\bibnamefont
  {Meschede}}, \bibinfo {author} {\bibfnamefont {H.}~\bibnamefont {Walther}},\
  and\ \bibinfo {author} {\bibfnamefont {G.}~\bibnamefont {M{\"{u}}ller}},\
  }\href {https://doi.org/10.1103/PhysRevLett.54.551} {\bibfield  {journal}
  {\bibinfo  {journal} {Phys. Rev. Lett.}\ }\textbf {\bibinfo {volume} {54}},\
  \bibinfo {pages} {551} (\bibinfo {year} {1985})}\BibitemShut {NoStop}%
\bibitem [{\citenamefont {Rempe}\ \emph {et~al.}(1987)\citenamefont {Rempe},
  \citenamefont {Walther},\ and\ \citenamefont {Klein}}]{REMPE1987}%
  \BibitemOpen
  \bibfield  {author} {\bibinfo {author} {\bibfnamefont {G.}~\bibnamefont
  {Rempe}}, \bibinfo {author} {\bibfnamefont {H.}~\bibnamefont {Walther}},\
  and\ \bibinfo {author} {\bibfnamefont {N.}~\bibnamefont {Klein}},\ }\href
  {https://doi.org/10.1103/PhysRevLett.58.353} {\bibfield  {journal} {\bibinfo
  {journal} {Phys. Rev. Lett.}\ }\textbf {\bibinfo {volume} {58}},\ \bibinfo
  {pages} {353} (\bibinfo {year} {1987})}\BibitemShut {NoStop}%
\bibitem [{\citenamefont {Brune}\ \emph {et~al.}(1996)\citenamefont {Brune},
  \citenamefont {Schmidt-Kaler}, \citenamefont {Maali}, \citenamefont {Dreyer},
  \citenamefont {Hagley}, \citenamefont {Raimond},\ and\ \citenamefont
  {Haroche}}]{Brune1996}%
  \BibitemOpen
  \bibfield  {author} {\bibinfo {author} {\bibfnamefont {M.}~\bibnamefont
  {Brune}}, \bibinfo {author} {\bibfnamefont {F.}~\bibnamefont
  {Schmidt-Kaler}}, \bibinfo {author} {\bibfnamefont {A.}~\bibnamefont
  {Maali}}, \bibinfo {author} {\bibfnamefont {J.}~\bibnamefont {Dreyer}},
  \bibinfo {author} {\bibfnamefont {E.}~\bibnamefont {Hagley}}, \bibinfo
  {author} {\bibfnamefont {J.~M.}\ \bibnamefont {Raimond}},\ and\ \bibinfo
  {author} {\bibfnamefont {S.}~\bibnamefont {Haroche}},\ }\href
  {https://doi.org/10.1103/PhysRevLett.76.1800} {\bibfield  {journal} {\bibinfo
   {journal} {Phys. Rev. Lett.}\ }\textbf {\bibinfo {volume} {76}},\ \bibinfo
  {pages} {1800} (\bibinfo {year} {1996})}\BibitemShut {NoStop}%
\bibitem [{\citenamefont {Kukli{\'{n}}ski}\ and\ \citenamefont
  {Madajczyk}(1988)}]{Kuklinski1988}%
  \BibitemOpen
  \bibfield  {author} {\bibinfo {author} {\bibfnamefont {J.~R.}\ \bibnamefont
  {Kukli{\'{n}}ski}}\ and\ \bibinfo {author} {\bibfnamefont {J.~L.}\
  \bibnamefont {Madajczyk}},\ }\href {https://doi.org/10.1103/PhysRevA.37.3175}
  {\bibfield  {journal} {\bibinfo  {journal} {Phys. Rev. A}\ }\textbf {\bibinfo
  {volume} {37}},\ \bibinfo {pages} {3175} (\bibinfo {year}
  {1988})}\BibitemShut {NoStop}%
\bibitem [{\citenamefont {Hillery}(1989)}]{Hillery1989}%
  \BibitemOpen
  \bibfield  {author} {\bibinfo {author} {\bibfnamefont {M.}~\bibnamefont
  {Hillery}},\ }\href {https://doi.org/10.1103/PhysRevA.39.1556} {\bibfield
  {journal} {\bibinfo  {journal} {Phys. Rev. A}\ }\textbf {\bibinfo {volume}
  {39}},\ \bibinfo {pages} {1556} (\bibinfo {year} {1989})}\BibitemShut
  {NoStop}%
\bibitem [{\citenamefont {Hillery}(1987)}]{Hillery1987}%
  \BibitemOpen
  \bibfield  {author} {\bibinfo {author} {\bibfnamefont {M.}~\bibnamefont
  {Hillery}},\ }\href {https://doi.org/10.1103/PhysRevA.35.4186} {\bibfield
  {journal} {\bibinfo  {journal} {Phys. Rev. A}\ }\textbf {\bibinfo {volume}
  {35}},\ \bibinfo {pages} {4186} (\bibinfo {year} {1987})}\BibitemShut
  {NoStop}%
\bibitem [{\citenamefont {Chumakov}\ \emph {et~al.}(1993)\citenamefont
  {Chumakov}, \citenamefont {Kozierowski},\ and\ \citenamefont
  {Sanchez-Mondragon}}]{Chumakov1993}%
  \BibitemOpen
  \bibfield  {author} {\bibinfo {author} {\bibfnamefont {S.~M.}\ \bibnamefont
  {Chumakov}}, \bibinfo {author} {\bibfnamefont {M.}~\bibnamefont
  {Kozierowski}},\ and\ \bibinfo {author} {\bibfnamefont {J.~J.}\ \bibnamefont
  {Sanchez-Mondragon}},\ }\href {https://doi.org/10.1103/PhysRevA.48.4594}
  {\bibfield  {journal} {\bibinfo  {journal} {Phys. Rev. A}\ }\textbf {\bibinfo
  {volume} {48}},\ \bibinfo {pages} {4594} (\bibinfo {year}
  {1993})}\BibitemShut {NoStop}%
\bibitem [{\citenamefont {Aravind}\ and\ \citenamefont
  {Hirschfelder}(1984)}]{Aravind1984}%
  \BibitemOpen
  \bibfield  {author} {\bibinfo {author} {\bibfnamefont {P.~K.}\ \bibnamefont
  {Aravind}}\ and\ \bibinfo {author} {\bibfnamefont {J.~O.}\ \bibnamefont
  {Hirschfelder}},\ }\href {https://doi.org/10.1021/j150665a002} {\bibfield
  {journal} {\bibinfo  {journal} {J. Phys. Chem.}\ }\textbf {\bibinfo {volume}
  {88}},\ \bibinfo {pages} {4788} (\bibinfo {year} {1984})}\BibitemShut
  {NoStop}%
\bibitem [{\citenamefont {Phoenix}\ and\ \citenamefont
  {Knight}(1988)}]{Phoenix1988}%
  \BibitemOpen
  \bibfield  {author} {\bibinfo {author} {\bibfnamefont {S.~J.~D.}\
  \bibnamefont {Phoenix}}\ and\ \bibinfo {author} {\bibfnamefont {P.~L.}\
  \bibnamefont {Knight}},\ }\href
  {https://doi.org/10.1016/0003-4916(88)90006-1} {\bibfield  {journal}
  {\bibinfo  {journal} {Ann. Phys.}\ }\textbf {\bibinfo {volume} {186}},\
  \bibinfo {pages} {381} (\bibinfo {year} {1988})}\BibitemShut {NoStop}%
\bibitem [{\citenamefont {Phoenix}\ and\ \citenamefont
  {Knight}(1991)}]{Phoenix1991}%
  \BibitemOpen
  \bibfield  {author} {\bibinfo {author} {\bibfnamefont {S.~J.~D.}\
  \bibnamefont {Phoenix}}\ and\ \bibinfo {author} {\bibfnamefont {P.~L.}\
  \bibnamefont {Knight}},\ }\href {https://doi.org/10.1103/PhysRevA.44.6023}
  {\bibfield  {journal} {\bibinfo  {journal} {Phys. Rev. A}\ }\textbf {\bibinfo
  {volume} {44}},\ \bibinfo {pages} {6023} (\bibinfo {year}
  {1991})}\BibitemShut {NoStop}%
\bibitem [{\citenamefont {Guo}\ and\ \citenamefont {Zheng}(1996)}]{Guo1996}%
  \BibitemOpen
  \bibfield  {author} {\bibinfo {author} {\bibfnamefont {G.-C.}\ \bibnamefont
  {Guo}}\ and\ \bibinfo {author} {\bibfnamefont {S.-B.}\ \bibnamefont
  {Zheng}},\ }\href {https://doi.org/10.1016/S0375-9601(96)00753-0} {\bibfield
  {journal} {\bibinfo  {journal} {Phys. Lett. A}\ }\textbf {\bibinfo {volume}
  {223}},\ \bibinfo {pages} {332} (\bibinfo {year} {1996})}\BibitemShut
  {NoStop}%
\bibitem [{\citenamefont {Gerry}\ and\ \citenamefont
  {Grobe}(1997)}]{Gerry1997}%
  \BibitemOpen
  \bibfield  {author} {\bibinfo {author} {\bibfnamefont {C.~C.}\ \bibnamefont
  {Gerry}}\ and\ \bibinfo {author} {\bibfnamefont {R.}~\bibnamefont {Grobe}},\
  }\href {https://doi.org/10.1103/PhysRevA.56.2390} {\bibfield  {journal}
  {\bibinfo  {journal} {Phys. Rev. A}\ }\textbf {\bibinfo {volume} {56}},\
  \bibinfo {pages} {2390} (\bibinfo {year} {1997})}\BibitemShut {NoStop}%
\bibitem [{\citenamefont {Blockley}\ \emph {et~al.}(1992)\citenamefont
  {Blockley}, \citenamefont {Walls},\ and\ \citenamefont
  {Risken}}]{Blockley1992}%
  \BibitemOpen
  \bibfield  {author} {\bibinfo {author} {\bibfnamefont {C.~A.}\ \bibnamefont
  {Blockley}}, \bibinfo {author} {\bibfnamefont {D.~F.}\ \bibnamefont
  {Walls}},\ and\ \bibinfo {author} {\bibfnamefont {H.}~\bibnamefont
  {Risken}},\ }\href {https://doi.org/10.1209/0295-5075/17/6/006} {\bibfield
  {journal} {\bibinfo  {journal} {EuroPhys. Lett. (EPL)}\ }\textbf {\bibinfo
  {volume} {17}},\ \bibinfo {pages} {509} (\bibinfo {year} {1992})}\BibitemShut
  {NoStop}%
\bibitem [{\citenamefont {Vogel}\ and\ \citenamefont {{de Matos
  Filho}}(1995)}]{Vogel1995}%
  \BibitemOpen
  \bibfield  {author} {\bibinfo {author} {\bibfnamefont {W.}~\bibnamefont
  {Vogel}}\ and\ \bibinfo {author} {\bibfnamefont {R.~L.}\ \bibnamefont {{de
  Matos Filho}}},\ }\href {https://doi.org/10.1103/PhysRevA.52.4214} {\bibfield
   {journal} {\bibinfo  {journal} {Phys. Rev. A}\ }\textbf {\bibinfo {volume}
  {52}},\ \bibinfo {pages} {4214} (\bibinfo {year} {1995})}\BibitemShut
  {NoStop}%
\bibitem [{\citenamefont {Pedernales}\ \emph {et~al.}(2015)\citenamefont
  {Pedernales}, \citenamefont {Lizuain}, \citenamefont {Felicetti},
  \citenamefont {Romero}, \citenamefont {Lamata},\ and\ \citenamefont
  {Solano}}]{Pedernales2015}%
  \BibitemOpen
  \bibfield  {author} {\bibinfo {author} {\bibfnamefont {J.~S.}\ \bibnamefont
  {Pedernales}}, \bibinfo {author} {\bibfnamefont {I.}~\bibnamefont {Lizuain}},
  \bibinfo {author} {\bibfnamefont {S.}~\bibnamefont {Felicetti}}, \bibinfo
  {author} {\bibfnamefont {G.}~\bibnamefont {Romero}}, \bibinfo {author}
  {\bibfnamefont {L.}~\bibnamefont {Lamata}},\ and\ \bibinfo {author}
  {\bibfnamefont {E.}~\bibnamefont {Solano}},\ }\href
  {https://doi.org/10.1038/srep15472} {\bibfield  {journal} {\bibinfo
  {journal} {Sci. Rep.}\ }\textbf {\bibinfo {volume} {5}},\ \bibinfo {pages}
  {15472} (\bibinfo {year} {2015})},\ \Eprint
  {https://arxiv.org/abs/1505.00698} {arXiv:1505.00698} \BibitemShut {NoStop}%
\bibitem [{\citenamefont {Bermudez}\ \emph {et~al.}(2007)\citenamefont
  {Bermudez}, \citenamefont {Martin-Delgado},\ and\ \citenamefont
  {Solano}}]{BERMUDEZ2007}%
  \BibitemOpen
  \bibfield  {author} {\bibinfo {author} {\bibfnamefont {A.}~\bibnamefont
  {Bermudez}}, \bibinfo {author} {\bibfnamefont {M.~A.}\ \bibnamefont
  {Martin-Delgado}},\ and\ \bibinfo {author} {\bibfnamefont {E.}~\bibnamefont
  {Solano}},\ }\href {https://doi.org/10.1103/PhysRevA.76.041801} {\bibfield
  {journal} {\bibinfo  {journal} {Phys. Rev. A}\ }\textbf {\bibinfo {volume}
  {76}},\ \bibinfo {pages} {041801} (\bibinfo {year} {2007})}\BibitemShut
  {NoStop}%
\bibitem [{\citenamefont {Larson}\ and\ \citenamefont
  {Mavrogordatos}(2021)}]{LARSON2021}%
  \BibitemOpen
  \bibfield  {author} {\bibinfo {author} {\bibfnamefont {J.}~\bibnamefont
  {Larson}}\ and\ \bibinfo {author} {\bibfnamefont {T.}~\bibnamefont
  {Mavrogordatos}},\ }\href {https://doi.org/10.1088/978-0-7503-3447-1} {\emph
  {\bibinfo {title} {{The Jaynes–Cummings Model and Its Descendants}}}}\
  (\bibinfo  {publisher} {IOP Publishing},\ \bibinfo {address} {Bristol},\
  \bibinfo {year} {2021})\BibitemShut {NoStop}%
\bibitem [{\citenamefont {Meher}\ and\ \citenamefont
  {Sivakumar}(2022)}]{Meher2022}%
  \BibitemOpen
  \bibfield  {author} {\bibinfo {author} {\bibfnamefont {N.}~\bibnamefont
  {Meher}}\ and\ \bibinfo {author} {\bibfnamefont {S.}~\bibnamefont
  {Sivakumar}},\ }\href {https://doi.org/10.1140/epjp/s13360-022-03172-x}
  {\bibfield  {journal} {\bibinfo  {journal} {Eur. Phys. J. Plus}\ }\textbf
  {\bibinfo {volume} {137}},\ \bibinfo {pages} {985} (\bibinfo {year}
  {2022})}\BibitemShut {NoStop}%
\bibitem [{\citenamefont {Lai}\ \emph {et~al.}(1991)\citenamefont {Lai},
  \citenamefont {Buek},\ and\ \citenamefont {Knight}}]{Lai1991}%
  \BibitemOpen
  \bibfield  {author} {\bibinfo {author} {\bibfnamefont {W.~K.}\ \bibnamefont
  {Lai}}, \bibinfo {author} {\bibfnamefont {V.}~\bibnamefont {Buek}},\ and\
  \bibinfo {author} {\bibfnamefont {P.~L.}\ \bibnamefont {Knight}},\ }\href
  {https://doi.org/10.1103/PhysRevA.44.6043} {\bibfield  {journal} {\bibinfo
  {journal} {Phys. Rev. A}\ }\textbf {\bibinfo {volume} {44}},\ \bibinfo
  {pages} {6043} (\bibinfo {year} {1991})}\BibitemShut {NoStop}%
\bibitem [{\citenamefont {Wu}\ and\ \citenamefont {Yang}(1997)}]{Wu1997}%
  \BibitemOpen
  \bibfield  {author} {\bibinfo {author} {\bibfnamefont {Y.}~\bibnamefont
  {Wu}}\ and\ \bibinfo {author} {\bibfnamefont {X.}~\bibnamefont {Yang}},\
  }\href {https://doi.org/10.1103/PhysRevA.56.2443} {\bibfield  {journal}
  {\bibinfo  {journal} {Phys. Rev. A}\ }\textbf {\bibinfo {volume} {56}},\
  \bibinfo {pages} {2443} (\bibinfo {year} {1997})}\BibitemShut {NoStop}%
\bibitem [{\citenamefont {Dicke}(1954)}]{Dicke1954}%
  \BibitemOpen
  \bibfield  {author} {\bibinfo {author} {\bibfnamefont {R.~H.}\ \bibnamefont
  {Dicke}},\ }\href {https://doi.org/10.1103/PhysRev.93.99} {\bibfield
  {journal} {\bibinfo  {journal} {Phys. Rev.}\ }\textbf {\bibinfo {volume}
  {93}},\ \bibinfo {pages} {99} (\bibinfo {year} {1954})}\BibitemShut {NoStop}%
\bibitem [{\citenamefont {Tavis}\ and\ \citenamefont
  {Cummings}(1968)}]{Tavis1968}%
  \BibitemOpen
  \bibfield  {author} {\bibinfo {author} {\bibfnamefont {M.}~\bibnamefont
  {Tavis}}\ and\ \bibinfo {author} {\bibfnamefont {F.~W.}\ \bibnamefont
  {Cummings}},\ }\href {https://doi.org/10.1103/PhysRev.170.379} {\bibfield
  {journal} {\bibinfo  {journal} {Phys. Rev.}\ }\textbf {\bibinfo {volume}
  {170}},\ \bibinfo {pages} {379} (\bibinfo {year} {1968})}\BibitemShut
  {NoStop}%
\bibitem [{\citenamefont {Chaichian}\ \emph {et~al.}(1990)\citenamefont
  {Chaichian}, \citenamefont {Ellinas},\ and\ \citenamefont
  {Kulish}}]{Chaichian1990}%
  \BibitemOpen
  \bibfield  {author} {\bibinfo {author} {\bibfnamefont {M.}~\bibnamefont
  {Chaichian}}, \bibinfo {author} {\bibfnamefont {D.}~\bibnamefont {Ellinas}},\
  and\ \bibinfo {author} {\bibfnamefont {P.}~\bibnamefont {Kulish}},\ }\href
  {https://doi.org/10.1103/PhysRevLett.65.980} {\bibfield  {journal} {\bibinfo
  {journal} {Phys. Rev. Lett.}\ }\textbf {\bibinfo {volume} {65}},\ \bibinfo
  {pages} {980} (\bibinfo {year} {1990})}\BibitemShut {NoStop}%
\bibitem [{\citenamefont {Dehghani}\ \emph {et~al.}(2016)\citenamefont
  {Dehghani}, \citenamefont {Mojaveri}, \citenamefont {Shirin},\ and\
  \citenamefont {Faseghandis}}]{Dehghani2016}%
  \BibitemOpen
  \bibfield  {author} {\bibinfo {author} {\bibfnamefont {A.}~\bibnamefont
  {Dehghani}}, \bibinfo {author} {\bibfnamefont {B.}~\bibnamefont {Mojaveri}},
  \bibinfo {author} {\bibfnamefont {S.}~\bibnamefont {Shirin}},\ and\ \bibinfo
  {author} {\bibfnamefont {S.~A.}\ \bibnamefont {Faseghandis}},\ }\href
  {https://doi.org/10.1038/srep38069} {\bibfield  {journal} {\bibinfo
  {journal} {Sci. Rep.}\ }\textbf {\bibinfo {volume} {6}},\ \bibinfo {pages}
  {38069} (\bibinfo {year} {2016})}\BibitemShut {NoStop}%
\bibitem [{\citenamefont {Uhdre}\ \emph {et~al.}(2022)\citenamefont {Uhdre},
  \citenamefont {Cius},\ and\ \citenamefont {Andrade}}]{UHDRE2022}%
  \BibitemOpen
  \bibfield  {author} {\bibinfo {author} {\bibfnamefont {G.~M.}\ \bibnamefont
  {Uhdre}}, \bibinfo {author} {\bibfnamefont {D.}~\bibnamefont {Cius}},\ and\
  \bibinfo {author} {\bibfnamefont {F.~M.}\ \bibnamefont {Andrade}},\ }\href
  {https://doi.org/10.1103/PhysRevA.105.013703} {\bibfield  {journal} {\bibinfo
   {journal} {Phys. Rev. A}\ }\textbf {\bibinfo {volume} {105}},\ \bibinfo
  {pages} {013703} (\bibinfo {year} {2022})}\BibitemShut {NoStop}%
\bibitem [{\citenamefont {Bu\ifmmode~\check{z}\else \v{z}\fi{}ek}\ and\
  \citenamefont {Jex}(1990)}]{Buzek1990}%
  \BibitemOpen
  \bibfield  {author} {\bibinfo {author} {\bibfnamefont {V.}~\bibnamefont
  {Bu\ifmmode~\check{z}\else \v{z}\fi{}ek}}\ and\ \bibinfo {author}
  {\bibfnamefont {I.}~\bibnamefont {Jex}},\ }\href
  {https://doi.org/10.1016/0030-4018(90)90340-Y} {\bibfield  {journal}
  {\bibinfo  {journal} {Opt. Commun.}\ }\textbf {\bibinfo {volume} {78}},\
  \bibinfo {pages} {425} (\bibinfo {year} {1990})}\BibitemShut {NoStop}%
\bibitem [{\citenamefont {Bu\ifmmode~\check{z}\else \v{z}\fi{}ek}\ and\
  \citenamefont {Jex}(1991)}]{Buzek1991}%
  \BibitemOpen
  \bibfield  {author} {\bibinfo {author} {\bibfnamefont {V.}~\bibnamefont
  {Bu\ifmmode~\check{z}\else \v{z}\fi{}ek}}\ and\ \bibinfo {author}
  {\bibfnamefont {I.}~\bibnamefont {Jex}},\ }\href
  {https://doi.org/10.1080/09500349114550961} {\bibfield  {journal} {\bibinfo
  {journal} {J. Mod. Opt.}\ }\textbf {\bibinfo {volume} {38}},\ \bibinfo
  {pages} {987} (\bibinfo {year} {1991})}\BibitemShut {NoStop}%
\bibitem [{\citenamefont {Bocanegra-Garay}\ \emph {et~al.}(2024)\citenamefont
  {Bocanegra-Garay}, \citenamefont {Hernández-Sánchez}, \citenamefont
  {Ramos-Prieto}, \citenamefont {Soto-Eguibar},\ and\ \citenamefont
  {Moya-Cessa}}]{Bocanegra-Garay2024}%
  \BibitemOpen
  \bibfield  {author} {\bibinfo {author} {\bibfnamefont {I.~A.}\ \bibnamefont
  {Bocanegra-Garay}}, \bibinfo {author} {\bibfnamefont {L.}~\bibnamefont
  {Hernández-Sánchez}}, \bibinfo {author} {\bibfnamefont {I.}~\bibnamefont
  {Ramos-Prieto}}, \bibinfo {author} {\bibfnamefont {F.}~\bibnamefont
  {Soto-Eguibar}},\ and\ \bibinfo {author} {\bibfnamefont {H.~M.}\ \bibnamefont
  {Moya-Cessa}},\ }\href {https://doi.org/10.21468/SciPostPhys.16.1.007}
  {\bibfield  {journal} {\bibinfo  {journal} {SciPost Phys.}\ }\textbf
  {\bibinfo {volume} {16}},\ \bibinfo {pages} {007} (\bibinfo {year}
  {2024})}\BibitemShut {NoStop}%
\bibitem [{\citenamefont {Vidiella-Barranco}\ \emph {et~al.}(2025)\citenamefont
  {Vidiella-Barranco}, \citenamefont {Magalhães~de Castro}, \citenamefont
  {Sergi}, \citenamefont {Roversi}, \citenamefont {Messina},\ and\
  \citenamefont {Migliore}}]{Vidiella-Barranco2025}%
  \BibitemOpen
  \bibfield  {author} {\bibinfo {author} {\bibfnamefont {A.}~\bibnamefont
  {Vidiella-Barranco}}, \bibinfo {author} {\bibfnamefont {A.~S.}\ \bibnamefont
  {Magalhães~de Castro}}, \bibinfo {author} {\bibfnamefont {A.}~\bibnamefont
  {Sergi}}, \bibinfo {author} {\bibfnamefont {J.~A.}\ \bibnamefont {Roversi}},
  \bibinfo {author} {\bibfnamefont {A.}~\bibnamefont {Messina}},\ and\ \bibinfo
  {author} {\bibfnamefont {A.}~\bibnamefont {Migliore}},\ }\href
  {https://doi.org/https://doi.org/10.1002/andp.202500148} {\bibfield
  {journal} {\bibinfo  {journal} {Annalen der Physik}\ }\textbf {\bibinfo
  {volume} {537}},\ \bibinfo {pages} {e00148} (\bibinfo {year}
  {2025})}\BibitemShut {NoStop}%
\bibitem [{\citenamefont {Puri}\ and\ \citenamefont
  {Agarwal}(1986)}]{Puri1986}%
  \BibitemOpen
  \bibfield  {author} {\bibinfo {author} {\bibfnamefont {R.~R.}\ \bibnamefont
  {Puri}}\ and\ \bibinfo {author} {\bibfnamefont {G.~S.}\ \bibnamefont
  {Agarwal}},\ }\href {https://doi.org/10.1103/PhysRevA.33.3610} {\bibfield
  {journal} {\bibinfo  {journal} {Phys. Rev. A}\ }\textbf {\bibinfo {volume}
  {33}},\ \bibinfo {pages} {3610} (\bibinfo {year} {1986})}\BibitemShut
  {NoStop}%
\bibitem [{\citenamefont {Barnett}\ and\ \citenamefont
  {Knight}(1986)}]{Barnett1986}%
  \BibitemOpen
  \bibfield  {author} {\bibinfo {author} {\bibfnamefont {S.~M.}\ \bibnamefont
  {Barnett}}\ and\ \bibinfo {author} {\bibfnamefont {P.~L.}\ \bibnamefont
  {Knight}},\ }\href {https://doi.org/10.1103/PhysRevA.33.2444} {\bibfield
  {journal} {\bibinfo  {journal} {Phys. Rev. A}\ }\textbf {\bibinfo {volume}
  {33}},\ \bibinfo {pages} {2444} (\bibinfo {year} {1986})}\BibitemShut
  {NoStop}%
\bibitem [{\citenamefont {Scala}\ \emph {et~al.}(2007)\citenamefont {Scala},
  \citenamefont {Militello}, \citenamefont {Messina}, \citenamefont {Piilo},\
  and\ \citenamefont {Maniscalco}}]{Scala2007}%
  \BibitemOpen
  \bibfield  {author} {\bibinfo {author} {\bibfnamefont {M.}~\bibnamefont
  {Scala}}, \bibinfo {author} {\bibfnamefont {B.}~\bibnamefont {Militello}},
  \bibinfo {author} {\bibfnamefont {A.}~\bibnamefont {Messina}}, \bibinfo
  {author} {\bibfnamefont {J.}~\bibnamefont {Piilo}},\ and\ \bibinfo {author}
  {\bibfnamefont {S.}~\bibnamefont {Maniscalco}},\ }\href
  {https://doi.org/10.1103/PhysRevA.75.013811} {\bibfield  {journal} {\bibinfo
  {journal} {Phys. Rev. A}\ }\textbf {\bibinfo {volume} {75}},\ \bibinfo
  {pages} {013811} (\bibinfo {year} {2007})}\BibitemShut {NoStop}%
\bibitem [{\citenamefont {Nussenzveig}\ \emph {et~al.}(1993)\citenamefont
  {Nussenzveig}, \citenamefont {Bernardot}, \citenamefont {Brune},
  \citenamefont {Hare}, \citenamefont {Raimond}, \citenamefont {Haroche},\ and\
  \citenamefont {Gawlik}}]{Nussenzveig1993}%
  \BibitemOpen
  \bibfield  {author} {\bibinfo {author} {\bibfnamefont {P.}~\bibnamefont
  {Nussenzveig}}, \bibinfo {author} {\bibfnamefont {F.}~\bibnamefont
  {Bernardot}}, \bibinfo {author} {\bibfnamefont {M.}~\bibnamefont {Brune}},
  \bibinfo {author} {\bibfnamefont {J.}~\bibnamefont {Hare}}, \bibinfo {author}
  {\bibfnamefont {J.~M.}\ \bibnamefont {Raimond}}, \bibinfo {author}
  {\bibfnamefont {S.}~\bibnamefont {Haroche}},\ and\ \bibinfo {author}
  {\bibfnamefont {W.}~\bibnamefont {Gawlik}},\ }\href
  {https://doi.org/10.1103/PhysRevA.48.3991} {\bibfield  {journal} {\bibinfo
  {journal} {Phys. Rev. A}\ }\textbf {\bibinfo {volume} {48}},\ \bibinfo
  {pages} {3991} (\bibinfo {year} {1993})}\BibitemShut {NoStop}%
\bibitem [{\citenamefont {Brune}\ \emph {et~al.}(1994)\citenamefont {Brune},
  \citenamefont {Nussenzveig}, \citenamefont {Schmidt-Kaler}, \citenamefont
  {Bernardot}, \citenamefont {Maali}, \citenamefont {Raimond},\ and\
  \citenamefont {Haroche}}]{Brune1994}%
  \BibitemOpen
  \bibfield  {author} {\bibinfo {author} {\bibfnamefont {M.}~\bibnamefont
  {Brune}}, \bibinfo {author} {\bibfnamefont {P.}~\bibnamefont {Nussenzveig}},
  \bibinfo {author} {\bibfnamefont {F.}~\bibnamefont {Schmidt-Kaler}}, \bibinfo
  {author} {\bibfnamefont {F.}~\bibnamefont {Bernardot}}, \bibinfo {author}
  {\bibfnamefont {A.}~\bibnamefont {Maali}}, \bibinfo {author} {\bibfnamefont
  {J.~M.}\ \bibnamefont {Raimond}},\ and\ \bibinfo {author} {\bibfnamefont
  {S.}~\bibnamefont {Haroche}},\ }\href
  {https://doi.org/10.1103/PhysRevLett.72.3339} {\bibfield  {journal} {\bibinfo
   {journal} {Phys. Rev. Lett.}\ }\textbf {\bibinfo {volume} {72}},\ \bibinfo
  {pages} {3339} (\bibinfo {year} {1994})}\BibitemShut {NoStop}%
\bibitem [{\citenamefont {Haroche}(2013)}]{Haroche2013}%
  \BibitemOpen
  \bibfield  {author} {\bibinfo {author} {\bibfnamefont {S.}~\bibnamefont
  {Haroche}},\ }\href {https://doi.org/10.1103/RevModPhys.85.1083} {\bibfield
  {journal} {\bibinfo  {journal} {Red. Mod. Phys.}\ }\textbf {\bibinfo {volume}
  {85}},\ \bibinfo {pages} {1083} (\bibinfo {year} {2013})}\BibitemShut
  {NoStop}%
\bibitem [{\citenamefont {Meekhof}\ \emph {et~al.}(1996)\citenamefont
  {Meekhof}, \citenamefont {Monroe}, \citenamefont {King}, \citenamefont
  {Itano},\ and\ \citenamefont {Wineland}}]{Meekhof1996}%
  \BibitemOpen
  \bibfield  {author} {\bibinfo {author} {\bibfnamefont {D.~M.}\ \bibnamefont
  {Meekhof}}, \bibinfo {author} {\bibfnamefont {C.}~\bibnamefont {Monroe}},
  \bibinfo {author} {\bibfnamefont {B.~E.}\ \bibnamefont {King}}, \bibinfo
  {author} {\bibfnamefont {W.~M.}\ \bibnamefont {Itano}},\ and\ \bibinfo
  {author} {\bibfnamefont {D.~J.}\ \bibnamefont {Wineland}},\ }\href
  {https://doi.org/10.1103/PhysRevLett.76.1796} {\bibfield  {journal} {\bibinfo
   {journal} {Phys. Rev. Lett.}\ }\textbf {\bibinfo {volume} {76}},\ \bibinfo
  {pages} {1796} (\bibinfo {year} {1996})}\BibitemShut {NoStop}%
\bibitem [{\citenamefont {Leibfried}\ \emph {et~al.}(2003)\citenamefont
  {Leibfried}, \citenamefont {Blatt}, \citenamefont {Monroe},\ and\
  \citenamefont {Wineland}}]{Leibfried2003}%
  \BibitemOpen
  \bibfield  {author} {\bibinfo {author} {\bibfnamefont {D.}~\bibnamefont
  {Leibfried}}, \bibinfo {author} {\bibfnamefont {R.}~\bibnamefont {Blatt}},
  \bibinfo {author} {\bibfnamefont {C.}~\bibnamefont {Monroe}},\ and\ \bibinfo
  {author} {\bibfnamefont {D.}~\bibnamefont {Wineland}},\ }\href
  {https://doi.org/10.1103/RevModPhys.75.281} {\bibfield  {journal} {\bibinfo
  {journal} {Rev. Mod. Phys.}\ }\textbf {\bibinfo {volume} {75}},\ \bibinfo
  {pages} {281} (\bibinfo {year} {2003})}\BibitemShut {NoStop}%
\bibitem [{\citenamefont {de~Castro}\ \emph {et~al.}(2023)\citenamefont
  {de~Castro}, \citenamefont {Grimaudo}, \citenamefont {Valenti}, \citenamefont
  {Migliore}, \citenamefont {Nakazato},\ and\ \citenamefont
  {Messina}}]{DeCastro2023}%
  \BibitemOpen
  \bibfield  {author} {\bibinfo {author} {\bibfnamefont {A.~S.~M.}\
  \bibnamefont {de~Castro}}, \bibinfo {author} {\bibfnamefont {R.}~\bibnamefont
  {Grimaudo}}, \bibinfo {author} {\bibfnamefont {D.}~\bibnamefont {Valenti}},
  \bibinfo {author} {\bibfnamefont {A.}~\bibnamefont {Migliore}}, \bibinfo
  {author} {\bibfnamefont {H.}~\bibnamefont {Nakazato}},\ and\ \bibinfo
  {author} {\bibfnamefont {A.}~\bibnamefont {Messina}},\ }\href
  {https://doi.org/10.1140/epjp/s13360-023-04375-6} {\bibfield  {journal}
  {\bibinfo  {journal} {Eur. Phys. J. Plus}\ }\textbf {\bibinfo {volume}
  {138}},\ \bibinfo {pages} {766} (\bibinfo {year} {2023})}\BibitemShut
  {NoStop}%
\bibitem [{\citenamefont {Schlicher}(1989)}]{Schlicher1989}%
  \BibitemOpen
  \bibfield  {author} {\bibinfo {author} {\bibfnamefont {R.~R.}\ \bibnamefont
  {Schlicher}},\ }\href {https://doi.org/10.1016/0030-4018(89)90276-9}
  {\bibfield  {journal} {\bibinfo  {journal} {Opt. Commun.}\ }\textbf {\bibinfo
  {volume} {70}},\ \bibinfo {pages} {97} (\bibinfo {year} {1989})}\BibitemShut
  {NoStop}%
\bibitem [{\citenamefont {Fang}(1998)}]{Fang1998}%
  \BibitemOpen
  \bibfield  {author} {\bibinfo {author} {\bibfnamefont {M.-F.}\ \bibnamefont
  {Fang}},\ }\href {https://doi.org/10.1016/S0378-4371(98)00234-9} {\bibfield
  {journal} {\bibinfo  {journal} {Physica A Stat. Mech. Appl.}\ }\textbf
  {\bibinfo {volume} {259}},\ \bibinfo {pages} {193} (\bibinfo {year}
  {1998})}\BibitemShut {NoStop}%
\bibitem [{\citenamefont {Prants}\ and\ \citenamefont
  {Yacoupova}(1992)}]{Prants1992}%
  \BibitemOpen
  \bibfield  {author} {\bibinfo {author} {\bibfnamefont {S.}~\bibnamefont
  {Prants}}\ and\ \bibinfo {author} {\bibfnamefont {L.}~\bibnamefont
  {Yacoupova}},\ }\href {https://doi.org/10.1080/09500349214550991} {\bibfield
  {journal} {\bibinfo  {journal} {J. Mod. Opt.}\ }\textbf {\bibinfo {volume}
  {39}},\ \bibinfo {pages} {961} (\bibinfo {year} {1992})}\BibitemShut
  {NoStop}%
\bibitem [{\citenamefont {Dasgupta}(1999)}]{Dasgupta1999}%
  \BibitemOpen
  \bibfield  {author} {\bibinfo {author} {\bibfnamefont {A.}~\bibnamefont
  {Dasgupta}},\ }\href {https://doi.org/10.1088/1464-4266/1/1/003} {\bibfield
  {journal} {\bibinfo  {journal} {J. Opt. B}\ }\textbf {\bibinfo {volume}
  {1}},\ \bibinfo {pages} {14} (\bibinfo {year} {1999})}\BibitemShut {NoStop}%
\bibitem [{\citenamefont {Joshi}(1995)}]{Joshi1995}%
  \BibitemOpen
  \bibfield  {author} {\bibinfo {author} {\bibfnamefont {A.}~\bibnamefont
  {Joshi}},\ }\href {https://doi.org/10.1080/713824334} {\bibfield  {journal}
  {\bibinfo  {journal} {J. Mod. Opt.}\ }\textbf {\bibinfo {volume} {42}},\
  \bibinfo {pages} {2561} (\bibinfo {year} {1995})}\BibitemShut {NoStop}%
\bibitem [{\citenamefont {Larson}\ and\ \citenamefont
  {Stenholm}(2003)}]{Larson2003}%
  \BibitemOpen
  \bibfield  {author} {\bibinfo {author} {\bibfnamefont {J.}~\bibnamefont
  {Larson}}\ and\ \bibinfo {author} {\bibfnamefont {S.}~\bibnamefont
  {Stenholm}},\ }\href {https://doi.org/10.1080/09500340308234580} {\bibfield
  {journal} {\bibinfo  {journal} {J. Mod. Opt.}\ }\textbf {\bibinfo {volume}
  {50}},\ \bibinfo {pages} {2705} (\bibinfo {year} {2003})}\BibitemShut
  {NoStop}%
\bibitem [{\citenamefont {Keeling}\ and\ \citenamefont
  {Gurarie}(2008)}]{Keeling2008}%
  \BibitemOpen
  \bibfield  {author} {\bibinfo {author} {\bibfnamefont {J.}~\bibnamefont
  {Keeling}}\ and\ \bibinfo {author} {\bibfnamefont {V.}~\bibnamefont
  {Gurarie}},\ }\href {https://doi.org/10.1103/PhysRevLett.101.033001}
  {\bibfield  {journal} {\bibinfo  {journal} {Phys. Rev. Lett.}\ }\textbf
  {\bibinfo {volume} {101}},\ \bibinfo {pages} {033001} (\bibinfo {year}
  {2008})}\BibitemShut {NoStop}%
\bibitem [{\citenamefont {Joshi}\ and\ \citenamefont
  {Lawande}(1993)}]{Joshi1993}%
  \BibitemOpen
  \bibfield  {author} {\bibinfo {author} {\bibfnamefont {A.}~\bibnamefont
  {Joshi}}\ and\ \bibinfo {author} {\bibfnamefont {S.~V.}\ \bibnamefont
  {Lawande}},\ }\href {https://doi.org/10.1103/PhysRevA.48.2276} {\bibfield
  {journal} {\bibinfo  {journal} {Phys. Rev. A}\ }\textbf {\bibinfo {volume}
  {48}},\ \bibinfo {pages} {2276} (\bibinfo {year} {1993})}\BibitemShut
  {NoStop}%
\bibitem [{\citenamefont {Maldonado-Mundo}\ \emph {et~al.}(2012)\citenamefont
  {Maldonado-Mundo}, \citenamefont {{\"{O}}hberg}, \citenamefont {Lovett},\
  and\ \citenamefont {Andersson}}]{Maldonado-Mundo2012}%
  \BibitemOpen
  \bibfield  {author} {\bibinfo {author} {\bibfnamefont {D.}~\bibnamefont
  {Maldonado-Mundo}}, \bibinfo {author} {\bibfnamefont {P.}~\bibnamefont
  {{\"{O}}hberg}}, \bibinfo {author} {\bibfnamefont {B.~W.}\ \bibnamefont
  {Lovett}},\ and\ \bibinfo {author} {\bibfnamefont {E.}~\bibnamefont
  {Andersson}},\ }\href {https://doi.org/10.1103/PhysRevA.86.042107} {\bibfield
   {journal} {\bibinfo  {journal} {Phys. Rev. A}\ }\textbf {\bibinfo {volume}
  {86}},\ \bibinfo {pages} {042107} (\bibinfo {year} {2012})}\BibitemShut
  {NoStop}%
\bibitem [{\citenamefont {Dong}\ and\ \citenamefont
  {Petersen}(2010)}]{Dong2010}%
  \BibitemOpen
  \bibfield  {author} {\bibinfo {author} {\bibfnamefont {D.}~\bibnamefont
  {Dong}}\ and\ \bibinfo {author} {\bibfnamefont {I.}~\bibnamefont
  {Petersen}},\ }\href {https://doi.org/10.1049/iet-cta.2009.0508} {\bibfield
  {journal} {\bibinfo  {journal} {IET Control Theory Appl.}\ }\textbf {\bibinfo
  {volume} {4}},\ \bibinfo {pages} {2651} (\bibinfo {year} {2010})}\BibitemShut
  {NoStop}%
\bibitem [{\citenamefont {Nielsen}\ and\ \citenamefont
  {Chuang}(2010)}]{Nielsen2010}%
  \BibitemOpen
  \bibfield  {author} {\bibinfo {author} {\bibfnamefont {M.~A.}\ \bibnamefont
  {Nielsen}}\ and\ \bibinfo {author} {\bibfnamefont {I.~L.}\ \bibnamefont
  {Chuang}},\ }\href@noop {} {\emph {\bibinfo {title} {{Quantum Computation and
  Quantum Information: 10th Anniversary Edition}}}}\ (\bibinfo  {publisher}
  {Cambridge University Press},\ \bibinfo {address} {Cambridge},\ \bibinfo
  {year} {2010})\BibitemShut {NoStop}%
\bibitem [{\citenamefont {Hern{\'{a}}ndez-S{\'{a}}nchez}\ \emph
  {et~al.}(2024)\citenamefont {Hern{\'{a}}ndez-S{\'{a}}nchez}, \citenamefont
  {Bocanegra-Garay}, \citenamefont {Ramos-Prieto}, \citenamefont
  {Soto-Eguibar},\ and\ \citenamefont {Moya-Cessa}}]{Hernandez-Sanchez2024}%
  \BibitemOpen
  \bibfield  {author} {\bibinfo {author} {\bibfnamefont {L.}~\bibnamefont
  {Hern{\'{a}}ndez-S{\'{a}}nchez}}, \bibinfo {author} {\bibfnamefont {I.~A.}\
  \bibnamefont {Bocanegra-Garay}}, \bibinfo {author} {\bibfnamefont
  {I.}~\bibnamefont {Ramos-Prieto}}, \bibinfo {author} {\bibfnamefont
  {F.}~\bibnamefont {Soto-Eguibar}},\ and\ \bibinfo {author} {\bibfnamefont
  {H.~M.}\ \bibnamefont {Moya-Cessa}},\ }\href
  {https://doi.org/10.1364/JOSAB.522587} {\bibfield  {journal} {\bibinfo
  {journal} {J. Opt. Soc. Am. B}\ }\textbf {\bibinfo {volume} {41}},\ \bibinfo
  {pages} {C68} (\bibinfo {year} {2024})}\BibitemShut {NoStop}%
\bibitem [{\citenamefont {Bose}\ \emph {et~al.}(2001)\citenamefont {Bose},
  \citenamefont {Fuentes-Guridi}, \citenamefont {Knight},\ and\ \citenamefont
  {Vedral}}]{Bose2001}%
  \BibitemOpen
  \bibfield  {author} {\bibinfo {author} {\bibfnamefont {S.}~\bibnamefont
  {Bose}}, \bibinfo {author} {\bibfnamefont {I.}~\bibnamefont
  {Fuentes-Guridi}}, \bibinfo {author} {\bibfnamefont {P.~L.}\ \bibnamefont
  {Knight}},\ and\ \bibinfo {author} {\bibfnamefont {V.}~\bibnamefont
  {Vedral}},\ }\href {https://doi.org/10.1103/PhysRevLett.87.050401} {\bibfield
   {journal} {\bibinfo  {journal} {Phys. Rev. Lett.}\ }\textbf {\bibinfo
  {volume} {87}},\ \bibinfo {pages} {050401} (\bibinfo {year}
  {2001})}\BibitemShut {NoStop}%
\bibitem [{\citenamefont {Yan}(2009)}]{Yan2009}%
  \BibitemOpen
  \bibfield  {author} {\bibinfo {author} {\bibfnamefont {X.-Q.}\ \bibnamefont
  {Yan}},\ }\href {https://doi.org/10.1016/j.chaos.2008.07.007} {\bibfield
  {journal} {\bibinfo  {journal} {Chaos Solit. Fractals}\ }\textbf {\bibinfo
  {volume} {41}},\ \bibinfo {pages} {1645} (\bibinfo {year}
  {2009})}\BibitemShut {NoStop}%
\bibitem [{\citenamefont {Joshi}\ and\ \citenamefont {Xiao}(2004)}]{Joshi2004}%
  \BibitemOpen
  \bibfield  {author} {\bibinfo {author} {\bibfnamefont {A.}~\bibnamefont
  {Joshi}}\ and\ \bibinfo {author} {\bibfnamefont {M.}~\bibnamefont {Xiao}},\
  }\href {https://doi.org/10.1364/JOSAB.21.001685} {\bibfield  {journal}
  {\bibinfo  {journal} {J. Opt. Soc. Am. B}\ }\textbf {\bibinfo {volume}
  {21}},\ \bibinfo {pages} {1685} (\bibinfo {year} {2004})}\BibitemShut
  {NoStop}%
\bibitem [{\citenamefont {Azuma}(2008)}]{Azuma2008}%
  \BibitemOpen
  \bibfield  {author} {\bibinfo {author} {\bibfnamefont {H.}~\bibnamefont
  {Azuma}},\ }\href {https://doi.org/10.1103/PhysRevA.77.063820} {\bibfield
  {journal} {\bibinfo  {journal} {Phys. Rev. A}\ }\textbf {\bibinfo {volume}
  {77}},\ \bibinfo {pages} {063820} (\bibinfo {year} {2008})}\BibitemShut
  {NoStop}%
\bibitem [{\citenamefont {Arancibia-Bulnes}\ \emph {et~al.}(1993)\citenamefont
  {Arancibia-Bulnes}, \citenamefont {Chumakov},\ and\ \citenamefont
  {S{\'{a}}nchez-Mondrag{\'{o}}n}}]{Arancibia-Bulnes1993}%
  \BibitemOpen
  \bibfield  {author} {\bibinfo {author} {\bibfnamefont {C.}~\bibnamefont
  {Arancibia-Bulnes}}, \bibinfo {author} {\bibfnamefont {S.}~\bibnamefont
  {Chumakov}},\ and\ \bibinfo {author} {\bibfnamefont {J.}~\bibnamefont
  {S{\'{a}}nchez-Mondrag{\'{o}}n}},\ }\href
  {https://doi.org/10.1080/09500349314552101} {\bibfield  {journal} {\bibinfo
  {journal} {J. Mod. Opt.}\ }\textbf {\bibinfo {volume} {40}},\ \bibinfo
  {pages} {2071} (\bibinfo {year} {1993})}\BibitemShut {NoStop}%
\bibitem [{\citenamefont {Azuma}\ and\ \citenamefont {Ban}(2014)}]{Azuma2014}%
  \BibitemOpen
  \bibfield  {author} {\bibinfo {author} {\bibfnamefont {H.}~\bibnamefont
  {Azuma}}\ and\ \bibinfo {author} {\bibfnamefont {M.}~\bibnamefont {Ban}},\
  }\href {https://doi.org/10.1016/j.physd.2014.04.009} {\bibfield  {journal}
  {\bibinfo  {journal} {Phys. D: Nonlinear Phenom.}\ }\textbf {\bibinfo
  {volume} {280-281}},\ \bibinfo {pages} {22} (\bibinfo {year}
  {2014})}\BibitemShut {NoStop}%
\bibitem [{\citenamefont {Azuma}\ and\ \citenamefont {Ban}(2015)}]{Azuma2015}%
  \BibitemOpen
  \bibfield  {author} {\bibinfo {author} {\bibfnamefont {H.}~\bibnamefont
  {Azuma}}\ and\ \bibinfo {author} {\bibfnamefont {M.}~\bibnamefont {Ban}},\
  }\href {https://doi.org/10.1016/j.physd.2015.05.006} {\bibfield  {journal}
  {\bibinfo  {journal} {Physica D: Nonlinear Phenomena}\ }\textbf {\bibinfo
  {volume} {308}},\ \bibinfo {pages} {127} (\bibinfo {year}
  {2015})}\BibitemShut {NoStop}%
\bibitem [{\citenamefont {Larson}(2007)}]{Larson2007}%
  \BibitemOpen
  \bibfield  {author} {\bibinfo {author} {\bibfnamefont {J.}~\bibnamefont
  {Larson}},\ }\href {https://doi.org/10.1088/0031-8949/76/2/007} {\bibfield
  {journal} {\bibinfo  {journal} {Phys. Scr.}\ }\textbf {\bibinfo {volume}
  {76}},\ \bibinfo {pages} {146} (\bibinfo {year} {2007})}\BibitemShut
  {NoStop}%
\bibitem [{\citenamefont {Scully}\ and\ \citenamefont
  {Zubairy}(1997)}]{Scully1997}%
  \BibitemOpen
  \bibfield  {author} {\bibinfo {author} {\bibfnamefont {M.~O.}\ \bibnamefont
  {Scully}}\ and\ \bibinfo {author} {\bibfnamefont {M.~S.}\ \bibnamefont
  {Zubairy}},\ }\href {https://doi.org/10.1017/CBO9780511813993} {\emph
  {\bibinfo {title} {{Quantum Optics}}}}\ (\bibinfo  {publisher} {Cambridge
  University Press},\ \bibinfo {address} {Cambridge},\ \bibinfo {year}
  {1997})\BibitemShut {NoStop}%
\bibitem [{\citenamefont {Gerry}\ and\ \citenamefont
  {Knight}(2004)}]{GERRY2005}%
  \BibitemOpen
  \bibfield  {author} {\bibinfo {author} {\bibfnamefont {C.}~\bibnamefont
  {Gerry}}\ and\ \bibinfo {author} {\bibfnamefont {P.}~\bibnamefont {Knight}},\
  }\href {https://doi.org/10.1017/CBO9780511791239} {\emph {\bibinfo {title}
  {{Introductory Quantum Optics}}}},\ \bibinfo {edition} {2nd}\ ed.\ (\bibinfo
  {publisher} {Cambridge University Press},\ \bibinfo {address} {Cambridge},\
  \bibinfo {year} {2004})\BibitemShut {NoStop}%
\bibitem [{\citenamefont {Braak}\ \emph {et~al.}(2016)\citenamefont {Braak},
  \citenamefont {Chen}, \citenamefont {Batchelor},\ and\ \citenamefont
  {Solano}}]{Braak2016}%
  \BibitemOpen
  \bibfield  {author} {\bibinfo {author} {\bibfnamefont {D.}~\bibnamefont
  {Braak}}, \bibinfo {author} {\bibfnamefont {Q.-H.}\ \bibnamefont {Chen}},
  \bibinfo {author} {\bibfnamefont {M.~T.}\ \bibnamefont {Batchelor}},\ and\
  \bibinfo {author} {\bibfnamefont {E.}~\bibnamefont {Solano}},\ }\href
  {https://doi.org/10.1088/1751-8113/49/30/300301} {\bibfield  {journal}
  {\bibinfo  {journal} {J. Phys. A}\ }\textbf {\bibinfo {volume} {49}},\
  \bibinfo {pages} {300301} (\bibinfo {year} {2016})}\BibitemShut {NoStop}%
\bibitem [{\citenamefont {Klimov}\ and\ \citenamefont
  {Chumakov}(2009)}]{Klimov2009}%
  \BibitemOpen
  \bibfield  {author} {\bibinfo {author} {\bibfnamefont {A.~B.}\ \bibnamefont
  {Klimov}}\ and\ \bibinfo {author} {\bibfnamefont {S.~M.}\ \bibnamefont
  {Chumakov}},\ }\href {https://doi.org/10.1002/9783527624003} {\emph {\bibinfo
  {title} {{A Group‐Theoretical Approach to Quantum Optics}}}}\ (\bibinfo
  {publisher} {Wiley-VCH},\ \bibinfo {address} {Weinheim},\ \bibinfo {year}
  {2009})\BibitemShut {NoStop}%
\bibitem [{\citenamefont {Cantuba}(2024)}]{Cantuba2024}%
  \BibitemOpen
  \bibfield  {author} {\bibinfo {author} {\bibfnamefont {R.~R.~S.}\
  \bibnamefont {Cantuba}},\ }\href {https://doi.org/10.24330/ieja.1326849}
  {\bibfield  {journal} {\bibinfo  {journal} {Int. Electron. J. Algebra}\
  }\textbf {\bibinfo {volume} {35}},\ \bibinfo {pages} {32} (\bibinfo {year}
  {2024})}\BibitemShut {NoStop}%
\bibitem [{\citenamefont {Kasper}\ \emph {et~al.}(2020)\citenamefont {Kasper},
  \citenamefont {Juzeliūnas}, \citenamefont {Lewenstein}, \citenamefont
  {Jendrzejewski},\ and\ \citenamefont {Zohar}}]{Kasper2020}%
  \BibitemOpen
  \bibfield  {author} {\bibinfo {author} {\bibfnamefont {V.}~\bibnamefont
  {Kasper}}, \bibinfo {author} {\bibfnamefont {G.}~\bibnamefont {Juzeliūnas}},
  \bibinfo {author} {\bibfnamefont {M.}~\bibnamefont {Lewenstein}}, \bibinfo
  {author} {\bibfnamefont {F.}~\bibnamefont {Jendrzejewski}},\ and\ \bibinfo
  {author} {\bibfnamefont {E.}~\bibnamefont {Zohar}},\ }\href
  {https://doi.org/10.1088/1367-2630/abb961} {\bibfield  {journal} {\bibinfo
  {journal} {New J. Phys}\ }\textbf {\bibinfo {volume} {22}},\ \bibinfo {pages}
  {103027} (\bibinfo {year} {2020})}\BibitemShut {NoStop}%
\bibitem [{\citenamefont {Bina}(2012)}]{Bina2012}%
  \BibitemOpen
  \bibfield  {author} {\bibinfo {author} {\bibfnamefont {M.}~\bibnamefont
  {Bina}},\ }\href {https://doi.org/10.1140/epjst/e2012-01541-3} {\bibfield
  {journal} {\bibinfo  {journal} {Eur. Phys. J. Special Topics}\ }\textbf
  {\bibinfo {volume} {203}},\ \bibinfo {pages} {163} (\bibinfo {year}
  {2012})}\BibitemShut {NoStop}%
\bibitem [{\citenamefont {Ju{\'{a}}rez-Amaro}\ \emph
  {et~al.}(2015)\citenamefont {Ju{\'{a}}rez-Amaro}, \citenamefont
  {Z{\'{u}}{\~{n}}iga-Segundo},\ and\ \citenamefont
  {Moya-Cessa}}]{Juarez-Amaro2015}%
  \BibitemOpen
  \bibfield  {author} {\bibinfo {author} {\bibfnamefont {R.}~\bibnamefont
  {Ju{\'{a}}rez-Amaro}}, \bibinfo {author} {\bibfnamefont {A.}~\bibnamefont
  {Z{\'{u}}{\~{n}}iga-Segundo}},\ and\ \bibinfo {author} {\bibfnamefont
  {H.~M.}\ \bibnamefont {Moya-Cessa}},\ }\href
  {https://doi.org/10.12785/amis/090136} {\bibfield  {journal} {\bibinfo
  {journal} {Appl. Math. Inf.}\ }\textbf {\bibinfo {volume} {9}},\ \bibinfo
  {pages} {299} (\bibinfo {year} {2015})}\BibitemShut {NoStop}%
\bibitem [{\citenamefont {Sakurai}\ and\ \citenamefont
  {Napolitano}(2020)}]{SAKURAI2020}%
  \BibitemOpen
  \bibfield  {author} {\bibinfo {author} {\bibfnamefont {J.~J.}\ \bibnamefont
  {Sakurai}}\ and\ \bibinfo {author} {\bibfnamefont {J.}~\bibnamefont
  {Napolitano}},\ }\href {https://doi.org/10.1017/9781108587280} {\emph
  {\bibinfo {title} {{Modern Quantum Mechanics}}}},\ \bibinfo {edition} {3rd}\
  ed.\ (\bibinfo  {publisher} {Cambridge University Press},\ \bibinfo {address}
  {Cambridge},\ \bibinfo {year} {2020})\BibitemShut {NoStop}%
\bibitem [{\citenamefont {Abramowitz}\ and\ \citenamefont
  {Stegun}(1972)}]{Abramowitz1972}%
  \BibitemOpen
  \bibinfo {editor} {\bibfnamefont {M.}~\bibnamefont {Abramowitz}}\ and\
  \bibinfo {editor} {\bibfnamefont {I.~A.}\ \bibnamefont {Stegun}},\ eds.,\
  \href@noop {} {\emph {\bibinfo {title} {{Handbook of Mathematical Functions
  with Formulas, Graphs and Mathematical Tables}}}},\ \bibinfo {edition} {1st}\
  ed.\ (\bibinfo  {publisher} {Dover},\ \bibinfo {year} {1972})\BibitemShut
  {NoStop}%
\bibitem [{\citenamefont {Glauber}(1963)}]{Glauber1963c}%
  \BibitemOpen
  \bibfield  {author} {\bibinfo {author} {\bibfnamefont {R.~J.}\ \bibnamefont
  {Glauber}},\ }\href {https://doi.org/10.1103/PhysRev.131.2766} {\bibfield
  {journal} {\bibinfo  {journal} {Phys. Rev.}\ }\textbf {\bibinfo {volume}
  {131}},\ \bibinfo {pages} {2766} (\bibinfo {year} {1963})}\BibitemShut
  {NoStop}%
\bibitem [{\citenamefont {Zhang}\ and\ \citenamefont {Jing}(2024)}]{Zhang2024}%
  \BibitemOpen
  \bibfield  {author} {\bibinfo {author} {\bibfnamefont {C.-Y.}\ \bibnamefont
  {Zhang}}\ and\ \bibinfo {author} {\bibfnamefont {J.}~\bibnamefont {Jing}},\
  }\href {https://doi.org/10.1103/PhysRevA.110.042421} {\bibfield  {journal}
  {\bibinfo  {journal} {Phys. Rev. A}\ }\textbf {\bibinfo {volume} {110}},\
  \bibinfo {pages} {042421} (\bibinfo {year} {2024})}\BibitemShut {NoStop}%
\bibitem [{\citenamefont {Haroche}\ and\ \citenamefont
  {Raimond}(2006)}]{Haroche2006}%
  \BibitemOpen
  \bibfield  {author} {\bibinfo {author} {\bibfnamefont {S.}~\bibnamefont
  {Haroche}}\ and\ \bibinfo {author} {\bibfnamefont {J.-M.}\ \bibnamefont
  {Raimond}},\ }\href
  {https://doi.org/10.1093/acprof:oso/9780198509141.001.0001} {\emph {\bibinfo
  {title} {{Exploring the Quantum}}}}\ (\bibinfo  {publisher} {Oxford
  University Press},\ \bibinfo {address} {Oxford},\ \bibinfo {year}
  {2006})\BibitemShut {NoStop}%
\bibitem [{\citenamefont {Arroyo-Correa}\ and\ \citenamefont
  {Sanchez-Mondragon}(1990)}]{Arroyo-Correa1990}%
  \BibitemOpen
  \bibfield  {author} {\bibinfo {author} {\bibfnamefont {G.}~\bibnamefont
  {Arroyo-Correa}}\ and\ \bibinfo {author} {\bibfnamefont {J.~J.}\ \bibnamefont
  {Sanchez-Mondragon}},\ }\href {https://doi.org/10.1088/0954-8998/2/6/001}
  {\bibfield  {journal} {\bibinfo  {journal} {Quant. Optics}\ }\textbf
  {\bibinfo {volume} {2}},\ \bibinfo {pages} {409} (\bibinfo {year}
  {1990})}\BibitemShut {NoStop}%
\bibitem [{\citenamefont {Horodecki}\ \emph {et~al.}(2009)\citenamefont
  {Horodecki}, \citenamefont {Horodecki}, \citenamefont {Horodecki},\ and\
  \citenamefont {Horodecki}}]{Horodecki2009}%
  \BibitemOpen
  \bibfield  {author} {\bibinfo {author} {\bibfnamefont {R.}~\bibnamefont
  {Horodecki}}, \bibinfo {author} {\bibfnamefont {P.}~\bibnamefont
  {Horodecki}}, \bibinfo {author} {\bibfnamefont {M.}~\bibnamefont
  {Horodecki}},\ and\ \bibinfo {author} {\bibfnamefont {K.}~\bibnamefont
  {Horodecki}},\ }\href {https://doi.org/10.1103/RevModPhys.81.865} {\bibfield
  {journal} {\bibinfo  {journal} {Rev. Mod. Phys.}\ }\textbf {\bibinfo {volume}
  {81}},\ \bibinfo {pages} {865} (\bibinfo {year} {2009})}\BibitemShut
  {NoStop}%
\bibitem [{\citenamefont {Schr\"odinger}(1935)}]{Schrodinger1935}%
  \BibitemOpen
  \bibfield  {author} {\bibinfo {author} {\bibfnamefont {E.}~\bibnamefont
  {Schr\"odinger}},\ }\href {https://doi.org/10.1007/BF01491891} {\bibfield
  {journal} {\bibinfo  {journal} {Naturwissenschaften}\ }\textbf {\bibinfo
  {volume} {23}},\ \bibinfo {pages} {807} (\bibinfo {year} {1935})}\BibitemShut
  {NoStop}%
\bibitem [{\citenamefont {Einstein}\ \emph {et~al.}(1935)\citenamefont
  {Einstein}, \citenamefont {Podolsky},\ and\ \citenamefont
  {Rosen}}]{Einstein1935}%
  \BibitemOpen
  \bibfield  {author} {\bibinfo {author} {\bibfnamefont {A.}~\bibnamefont
  {Einstein}}, \bibinfo {author} {\bibfnamefont {B.}~\bibnamefont {Podolsky}},\
  and\ \bibinfo {author} {\bibfnamefont {N.}~\bibnamefont {Rosen}},\ }\href
  {https://doi.org/10.1103/PhysRev.47.777} {\bibfield  {journal} {\bibinfo
  {journal} {Phys. Rev.}\ }\textbf {\bibinfo {volume} {47}},\ \bibinfo {pages}
  {777} (\bibinfo {year} {1935})}\BibitemShut {NoStop}%
\bibitem [{\citenamefont {Bell}(1964)}]{Bell1964}%
  \BibitemOpen
  \bibfield  {author} {\bibinfo {author} {\bibfnamefont {J.~S.}\ \bibnamefont
  {Bell}},\ }\href {https://doi.org/10.1103/PhysicsPhysiqueFizika.1.195}
  {\bibfield  {journal} {\bibinfo  {journal} {Physics Physique Fizika}\
  }\textbf {\bibinfo {volume} {1}},\ \bibinfo {pages} {195} (\bibinfo {year}
  {1964})}\BibitemShut {NoStop}%
\bibitem [{\citenamefont {von Neumann}(1927)}]{vonNeumann1927b}%
  \BibitemOpen
  \bibfield  {author} {\bibinfo {author} {\bibfnamefont {J.}~\bibnamefont {von
  Neumann}},\ }\href@noop {} {\bibfield  {journal} {\bibinfo  {journal} {Gott.
  Nachr. Math. Phys. Klass}\ } (\bibinfo {year} {1927})}\BibitemShut {NoStop}%
\bibitem [{\citenamefont {Phoenix}\ and\ \citenamefont
  {Knight}(1990)}]{Phoenix1990}%
  \BibitemOpen
  \bibfield  {author} {\bibinfo {author} {\bibfnamefont {S.~J.~D.}\
  \bibnamefont {Phoenix}}\ and\ \bibinfo {author} {\bibfnamefont {P.~L.}\
  \bibnamefont {Knight}},\ }\href {https://doi.org/10.1364/JOSAB.7.000116}
  {\bibfield  {journal} {\bibinfo  {journal} {J. Opt. Soc. Am. B}\ }\textbf
  {\bibinfo {volume} {7}},\ \bibinfo {pages} {116} (\bibinfo {year}
  {1990})}\BibitemShut {NoStop}%
\bibitem [{\citenamefont {Schmidt}(1907)}]{Schmidt1907}%
  \BibitemOpen
  \bibfield  {author} {\bibinfo {author} {\bibfnamefont {E.}~\bibnamefont
  {Schmidt}},\ }\href {https://doi.org/10.1007/BF01449770} {\bibfield
  {journal} {\bibinfo  {journal} {Math. Ann.}\ }\textbf {\bibinfo {volume}
  {63}},\ \bibinfo {pages} {433} (\bibinfo {year} {1907})}\BibitemShut
  {NoStop}%
\bibitem [{\citenamefont {Ekert}\ and\ \citenamefont
  {Knight}(1995)}]{Ekert1995}%
  \BibitemOpen
  \bibfield  {author} {\bibinfo {author} {\bibfnamefont {A.}~\bibnamefont
  {Ekert}}\ and\ \bibinfo {author} {\bibfnamefont {P.~L.}\ \bibnamefont
  {Knight}},\ }\href {https://doi.org/10.1119/1.17904} {\bibfield  {journal}
  {\bibinfo  {journal} {Am. J. Phys.}\ }\textbf {\bibinfo {volume} {63}},\
  \bibinfo {pages} {415} (\bibinfo {year} {1995})}\BibitemShut {NoStop}%
\bibitem [{\citenamefont {Araki}\ and\ \citenamefont {Lieb}(2002)}]{Araki2002}%
  \BibitemOpen
  \bibfield  {author} {\bibinfo {author} {\bibfnamefont {H.}~\bibnamefont
  {Araki}}\ and\ \bibinfo {author} {\bibfnamefont {E.~H.}\ \bibnamefont
  {Lieb}},\ }in\ \href {https://doi.org/10.1007/978-3-642-55925-9_4} {\emph
  {\bibinfo {booktitle} {Inequalities}}}\ (\bibinfo  {publisher} {Springer
  Berlin Heidelberg},\ \bibinfo {address} {Berlin, Heidelberg},\ \bibinfo
  {year} {2002})\ pp.\ \bibinfo {pages} {47--57}\BibitemShut {NoStop}%
\bibitem [{\citenamefont {Boukobza}\ and\ \citenamefont
  {Tannor}(2005)}]{Boukobza2005}%
  \BibitemOpen
  \bibfield  {author} {\bibinfo {author} {\bibfnamefont {E.}~\bibnamefont
  {Boukobza}}\ and\ \bibinfo {author} {\bibfnamefont {D.~J.}\ \bibnamefont
  {Tannor}},\ }\href {https://doi.org/10.1103/PhysRevA.71.063821} {\bibfield
  {journal} {\bibinfo  {journal} {Phys. Rev. A}\ }\textbf {\bibinfo {volume}
  {71}},\ \bibinfo {pages} {063821} (\bibinfo {year} {2005})}\BibitemShut
  {NoStop}%
\bibitem [{\citenamefont {Fasihi}\ and\ \citenamefont
  {Mojaveri}(2019)}]{Fasihi2019}%
  \BibitemOpen
  \bibfield  {author} {\bibinfo {author} {\bibfnamefont {M.~A.}\ \bibnamefont
  {Fasihi}}\ and\ \bibinfo {author} {\bibfnamefont {B.}~\bibnamefont
  {Mojaveri}},\ }\href {https://doi.org/10.1007/s11128-019-2195-8} {\bibfield
  {journal} {\bibinfo  {journal} {Quantum Inf. Process.}\ }\textbf {\bibinfo
  {volume} {18}},\ \bibinfo {pages} {75} (\bibinfo {year} {2019})}\BibitemShut
  {NoStop}%
\bibitem [{\citenamefont {Y{\"{o}}na{\c{c}}}\ \emph {et~al.}(2006)\citenamefont
  {Y{\"{o}}na{\c{c}}}, \citenamefont {Yu},\ and\ \citenamefont
  {Eberly}}]{Yonac2006}%
  \BibitemOpen
  \bibfield  {author} {\bibinfo {author} {\bibfnamefont {M.}~\bibnamefont
  {Y{\"{o}}na{\c{c}}}}, \bibinfo {author} {\bibfnamefont {T.}~\bibnamefont
  {Yu}},\ and\ \bibinfo {author} {\bibfnamefont {J.~H.}\ \bibnamefont
  {Eberly}},\ }\href {https://doi.org/10.1088/0953-4075/39/15/S09} {\bibfield
  {journal} {\bibinfo  {journal} {J. Phys. B}\ }\textbf {\bibinfo {volume}
  {39}},\ \bibinfo {pages} {S621} (\bibinfo {year} {2006})}\BibitemShut
  {NoStop}%
\bibitem [{\citenamefont {Prants}\ \emph {et~al.}(2006)\citenamefont {Prants},
  \citenamefont {Uleysky},\ and\ \citenamefont {Argonov}}]{Prants2006}%
  \BibitemOpen
  \bibfield  {author} {\bibinfo {author} {\bibfnamefont {S.~V.}\ \bibnamefont
  {Prants}}, \bibinfo {author} {\bibfnamefont {M.~Y.}\ \bibnamefont
  {Uleysky}},\ and\ \bibinfo {author} {\bibfnamefont {V.~Y.}\ \bibnamefont
  {Argonov}},\ }\href {https://doi.org/10.1103/PhysRevA.73.023807} {\bibfield
  {journal} {\bibinfo  {journal} {Phys. Rev. A}\ }\textbf {\bibinfo {volume}
  {73}},\ \bibinfo {pages} {023807} (\bibinfo {year} {2006})}\BibitemShut
  {NoStop}%
\bibitem [{\citenamefont {Tan}\ \emph {et~al.}(2011)\citenamefont {Tan},
  \citenamefont {Zhang},\ and\ \citenamefont {Zhu}}]{Tan2011}%
  \BibitemOpen
  \bibfield  {author} {\bibinfo {author} {\bibfnamefont {L.}~\bibnamefont
  {Tan}}, \bibinfo {author} {\bibfnamefont {Y.-Q.}\ \bibnamefont {Zhang}},\
  and\ \bibinfo {author} {\bibfnamefont {Z.-H.}\ \bibnamefont {Zhu}},\ }\href
  {https://doi.org/10.1088/1674-1056/20/7/070303} {\bibfield  {journal}
  {\bibinfo  {journal} {Chin. Phys. B}\ }\textbf {\bibinfo {volume} {20}},\
  \bibinfo {pages} {070303} (\bibinfo {year} {2011})}\BibitemShut {NoStop}%
\bibitem [{\citenamefont {Cheng}\ \emph {et~al.}(2018)\citenamefont {Cheng},
  \citenamefont {Chen},\ and\ \citenamefont {Shan}}]{Cheng2018}%
  \BibitemOpen
  \bibfield  {author} {\bibinfo {author} {\bibfnamefont {J.}~\bibnamefont
  {Cheng}}, \bibinfo {author} {\bibfnamefont {X.}~\bibnamefont {Chen}},\ and\
  \bibinfo {author} {\bibfnamefont {C.-J.}\ \bibnamefont {Shan}},\ }\href
  {https://doi.org/10.1007/s10773-018-3707-5} {\bibfield  {journal} {\bibinfo
  {journal} {Int. J. Theor. Phys.}\ }\textbf {\bibinfo {volume} {57}},\
  \bibinfo {pages} {1823} (\bibinfo {year} {2018})}\BibitemShut {NoStop}%
\bibitem [{\citenamefont {Bartzis}(1992)}]{Bartzis1992}%
  \BibitemOpen
  \bibfield  {author} {\bibinfo {author} {\bibfnamefont {V.}~\bibnamefont
  {Bartzis}},\ }\href {https://doi.org/10.1016/0378-4371(92)90399-B} {\bibfield
   {journal} {\bibinfo  {journal} {Phys. A: Stat. Mech. Appl.}\ }\textbf
  {\bibinfo {volume} {180}},\ \bibinfo {pages} {428} (\bibinfo {year}
  {1992})}\BibitemShut {NoStop}%
\bibitem [{\citenamefont {Yang}\ \emph {et~al.}(2006)\citenamefont {Yang},
  \citenamefont {Ya-Ping},\ and\ \citenamefont {Hong}}]{Yang2006}%
  \BibitemOpen
  \bibfield  {author} {\bibinfo {author} {\bibfnamefont {S.}~\bibnamefont
  {Yang}}, \bibinfo {author} {\bibfnamefont {Y.}~\bibnamefont {Ya-Ping}},\ and\
  \bibinfo {author} {\bibfnamefont {C.}~\bibnamefont {Hong}},\ }\href
  {https://doi.org/10.1088/0256-307X/23/5/020} {\bibfield  {journal} {\bibinfo
  {journal} {Chin. Phys. Lett.}\ }\textbf {\bibinfo {volume} {23}},\ \bibinfo
  {pages} {1136} (\bibinfo {year} {2006})}\BibitemShut {NoStop}%
\bibitem [{\citenamefont {Hu}\ and\ \citenamefont {Tan}(2014)}]{Hu2014}%
  \BibitemOpen
  \bibfield  {author} {\bibinfo {author} {\bibfnamefont {Y.-H.}\ \bibnamefont
  {Hu}}\ and\ \bibinfo {author} {\bibfnamefont {Y.-G.}\ \bibnamefont {Tan}},\
  }\href {https://doi.org/10.1088/0031-8949/89/7/075103} {\bibfield  {journal}
  {\bibinfo  {journal} {Phys. Scr.}\ }\textbf {\bibinfo {volume} {89}},\
  \bibinfo {pages} {075103} (\bibinfo {year} {2014})}\BibitemShut {NoStop}%
\bibitem [{\citenamefont {Wilkens}\ and\ \citenamefont
  {Meystre}(1992)}]{Wilkens1992}%
  \BibitemOpen
  \bibfield  {author} {\bibinfo {author} {\bibfnamefont {M.}~\bibnamefont
  {Wilkens}}\ and\ \bibinfo {author} {\bibfnamefont {P.}~\bibnamefont
  {Meystre}},\ }\href {https://doi.org/10.1016/0030-4018(92)90407-I} {\bibfield
   {journal} {\bibinfo  {journal} {Opt. Commun.}\ }\textbf {\bibinfo {volume}
  {94}},\ \bibinfo {pages} {66} (\bibinfo {year} {1992})}\BibitemShut {NoStop}%
\bibitem [{\citenamefont {Palma}\ and\ \citenamefont
  {Persico}(1992)}]{Palma1992}%
  \BibitemOpen
  \bibfield  {author} {\bibinfo {author} {\bibfnamefont {G.~M.}\ \bibnamefont
  {Palma}}\ and\ \bibinfo {author} {\bibfnamefont {F.~S.}\ \bibnamefont
  {Persico}},\ }\href {https://doi.org/10.1209/0295-5075/17/3/004} {\bibfield
  {journal} {\bibinfo  {journal} {EuroPhys. Lett. (EPL)}\ }\textbf {\bibinfo
  {volume} {17}},\ \bibinfo {pages} {207} (\bibinfo {year} {1992})}\BibitemShut
  {NoStop}%
\bibitem [{\citenamefont {Ren}\ \emph {et~al.}(1992)\citenamefont {Ren},
  \citenamefont {Cresser},\ and\ \citenamefont {Carmichael}}]{Ren1992}%
  \BibitemOpen
  \bibfield  {author} {\bibinfo {author} {\bibfnamefont {W.}~\bibnamefont
  {Ren}}, \bibinfo {author} {\bibfnamefont {J.~D.}\ \bibnamefont {Cresser}},\
  and\ \bibinfo {author} {\bibfnamefont {H.~J.}\ \bibnamefont {Carmichael}},\
  }\href {https://doi.org/10.1103/PhysRevA.46.7162} {\bibfield  {journal}
  {\bibinfo  {journal} {Phys. Rev. A}\ }\textbf {\bibinfo {volume} {46}},\
  \bibinfo {pages} {7162} (\bibinfo {year} {1992})}\BibitemShut {NoStop}%
\bibitem [{\citenamefont {Xie}\ \emph {et~al.}(2009)\citenamefont {Xie},
  \citenamefont {Jia},\ and\ \citenamefont {Yang}}]{Xie2009}%
  \BibitemOpen
  \bibfield  {author} {\bibinfo {author} {\bibfnamefont {S.}~\bibnamefont
  {Xie}}, \bibinfo {author} {\bibfnamefont {F.}~\bibnamefont {Jia}},\ and\
  \bibinfo {author} {\bibfnamefont {Y.}~\bibnamefont {Yang}},\ }\href
  {https://doi.org/10.1016/j.optcom.2009.03.024} {\bibfield  {journal}
  {\bibinfo  {journal} {Opt. Commun.}\ }\textbf {\bibinfo {volume} {282}},\
  \bibinfo {pages} {2642} (\bibinfo {year} {2009})}\BibitemShut {NoStop}%
\bibitem [{\citenamefont {Abdalla}\ \emph {et~al.}(2003)\citenamefont
  {Abdalla}, \citenamefont {Abdel-Aty},\ and\ \citenamefont
  {Obada}}]{Abdalla2003}%
  \BibitemOpen
  \bibfield  {author} {\bibinfo {author} {\bibfnamefont {M.}~\bibnamefont
  {Abdalla}}, \bibinfo {author} {\bibfnamefont {M.}~\bibnamefont {Abdel-Aty}},\
  and\ \bibinfo {author} {\bibfnamefont {A.-S.}\ \bibnamefont {Obada}},\ }\href
  {https://doi.org/10.1016/S0378-4371(03)00290-5} {\bibfield  {journal}
  {\bibinfo  {journal} {Phys. A: Stat. Mech. Appl.}\ }\textbf {\bibinfo
  {volume} {326}},\ \bibinfo {pages} {203} (\bibinfo {year}
  {2003})}\BibitemShut {NoStop}%
\bibitem [{\citenamefont {Abdel-Khalek}\ \emph {et~al.}(2015)\citenamefont
  {Abdel-Khalek}, \citenamefont {Quthami},\ and\ \citenamefont
  {Ahmed}}]{Abdel-Khalek2015}%
  \BibitemOpen
  \bibfield  {author} {\bibinfo {author} {\bibfnamefont {S.}~\bibnamefont
  {Abdel-Khalek}}, \bibinfo {author} {\bibfnamefont {M.}~\bibnamefont
  {Quthami}},\ and\ \bibinfo {author} {\bibfnamefont {M.~M.}\ \bibnamefont
  {Ahmed}},\ }\href {https://doi.org/10.1007/s10043-015-0044-2} {\bibfield
  {journal} {\bibinfo  {journal} {Opt. Rev.}\ }\textbf {\bibinfo {volume}
  {22}},\ \bibinfo {pages} {25} (\bibinfo {year} {2015})}\BibitemShut {NoStop}%
\bibitem [{\citenamefont {Haroche}\ and\ \citenamefont
  {Kleppner}(1989)}]{Haroche1989}%
  \BibitemOpen
  \bibfield  {author} {\bibinfo {author} {\bibfnamefont {S.}~\bibnamefont
  {Haroche}}\ and\ \bibinfo {author} {\bibfnamefont {D.}~\bibnamefont
  {Kleppner}},\ }\href {https://doi.org/10.1063/1.881201} {\bibfield  {journal}
  {\bibinfo  {journal} {Phys. Today}\ }\textbf {\bibinfo {volume} {42}},\
  \bibinfo {pages} {24} (\bibinfo {year} {1989})}\BibitemShut {NoStop}%
\bibitem [{\citenamefont {Raimond}\ \emph {et~al.}(2001)\citenamefont
  {Raimond}, \citenamefont {Brune},\ and\ \citenamefont
  {Haroche}}]{Raimond2001}%
  \BibitemOpen
  \bibfield  {author} {\bibinfo {author} {\bibfnamefont {J.~M.}\ \bibnamefont
  {Raimond}}, \bibinfo {author} {\bibfnamefont {M.}~\bibnamefont {Brune}},\
  and\ \bibinfo {author} {\bibfnamefont {S.}~\bibnamefont {Haroche}},\ }\href
  {https://doi.org/10.1103/RevModPhys.73.565} {\bibfield  {journal} {\bibinfo
  {journal} {Rev. Mod. Phys.}\ }\textbf {\bibinfo {volume} {73}},\ \bibinfo
  {pages} {565} (\bibinfo {year} {2001})}\BibitemShut {NoStop}%
\bibitem [{\citenamefont {Knight}\ and\ \citenamefont
  {Radmore}(1982)}]{Knight1982}%
  \BibitemOpen
  \bibfield  {author} {\bibinfo {author} {\bibfnamefont {P.}~\bibnamefont
  {Knight}}\ and\ \bibinfo {author} {\bibfnamefont {P.}~\bibnamefont
  {Radmore}},\ }\href {https://doi.org/10.1016/0375-9601(82)90625-9} {\bibfield
   {journal} {\bibinfo  {journal} {Phys. Lett. A}\ }\textbf {\bibinfo {volume}
  {90}},\ \bibinfo {pages} {342} (\bibinfo {year} {1982})}\BibitemShut
  {NoStop}%
\bibitem [{\citenamefont {von Foerster}(1975)}]{Foerster1975}%
  \BibitemOpen
  \bibfield  {author} {\bibinfo {author} {\bibfnamefont {T.}~\bibnamefont {von
  Foerster}},\ }\href {https://doi.org/10.1088/0305-4470/8/1/015} {\bibfield
  {journal} {\bibinfo  {journal} {J. Phys. A}\ }\textbf {\bibinfo {volume}
  {8}},\ \bibinfo {pages} {95} (\bibinfo {year} {1975})}\BibitemShut {NoStop}%
\bibitem [{\citenamefont {Buck}\ and\ \citenamefont
  {Sukumar}(1981)}]{Buck1981}%
  \BibitemOpen
  \bibfield  {author} {\bibinfo {author} {\bibfnamefont {B.}~\bibnamefont
  {Buck}}\ and\ \bibinfo {author} {\bibfnamefont {C.}~\bibnamefont {Sukumar}},\
  }\href {https://doi.org/10.1016/0375-9601(81)90042-6} {\bibfield  {journal}
  {\bibinfo  {journal} {Phys. Lett. A}\ }\textbf {\bibinfo {volume} {81}},\
  \bibinfo {pages} {132} (\bibinfo {year} {1981})}\BibitemShut {NoStop}%
\bibitem [{\citenamefont {Sukumar}\ and\ \citenamefont
  {Buck}(1981)}]{Sukumar1981}%
  \BibitemOpen
  \bibfield  {author} {\bibinfo {author} {\bibfnamefont {C.}~\bibnamefont
  {Sukumar}}\ and\ \bibinfo {author} {\bibfnamefont {B.}~\bibnamefont {Buck}},\
  }\href {https://doi.org/10.1016/0375-9601(81)90825-2} {\bibfield  {journal}
  {\bibinfo  {journal} {Phys. Lett. A}\ }\textbf {\bibinfo {volume} {83}},\
  \bibinfo {pages} {211} (\bibinfo {year} {1981})}\BibitemShut {NoStop}%
\bibitem [{\citenamefont {Klimov}\ and\ \citenamefont
  {Chumakov}(1999)}]{Klimov1999}%
  \BibitemOpen
  \bibfield  {author} {\bibinfo {author} {\bibfnamefont {A.}~\bibnamefont
  {Klimov}}\ and\ \bibinfo {author} {\bibfnamefont {S.}~\bibnamefont
  {Chumakov}},\ }\href {https://doi.org/10.1016/S0375-9601(99)00806-3}
  {\bibfield  {journal} {\bibinfo  {journal} {Phys. Lett. A}\ }\textbf
  {\bibinfo {volume} {264}},\ \bibinfo {pages} {100} (\bibinfo {year}
  {1999})}\BibitemShut {NoStop}%
\bibitem [{\citenamefont {Scheel}\ \emph {et~al.}(2003)\citenamefont {Scheel},
  \citenamefont {Eisert}, \citenamefont {Knight},\ and\ \citenamefont
  {Plenio}}]{Scheel2003}%
  \BibitemOpen
  \bibfield  {author} {\bibinfo {author} {\bibfnamefont {S.}~\bibnamefont
  {Scheel}}, \bibinfo {author} {\bibfnamefont {J.}~\bibnamefont {Eisert}},
  \bibinfo {author} {\bibfnamefont {P.~L.}\ \bibnamefont {Knight}},\ and\
  \bibinfo {author} {\bibfnamefont {M.~B.}\ \bibnamefont {Plenio}},\ }\href
  {https://doi.org/10.1080/09500340308234538} {\bibfield  {journal} {\bibinfo
  {journal} {J. Mod. Opt.}\ }\textbf {\bibinfo {volume} {50}},\ \bibinfo
  {pages} {881} (\bibinfo {year} {2003})}\BibitemShut {NoStop}%
\bibitem [{\citenamefont {Gea-Banacloche}(1992)}]{Gea-Banacloche1992}%
  \BibitemOpen
  \bibfield  {author} {\bibinfo {author} {\bibfnamefont {J.}~\bibnamefont
  {Gea-Banacloche}},\ }\href {https://doi.org/10.1016/0030-4018(92)90082-3}
  {\bibfield  {journal} {\bibinfo  {journal} {Opt. Commun.}\ }\textbf {\bibinfo
  {volume} {88}},\ \bibinfo {pages} {531} (\bibinfo {year} {1992})}\BibitemShut
  {NoStop}%
\bibitem [{\citenamefont {Zaheer}\ and\ \citenamefont
  {Zubairy}(1989)}]{Zaheer89}%
  \BibitemOpen
  \bibfield  {author} {\bibinfo {author} {\bibfnamefont {K.}~\bibnamefont
  {Zaheer}}\ and\ \bibinfo {author} {\bibfnamefont {M.~S.}\ \bibnamefont
  {Zubairy}},\ }\href {https://doi.org/10.1103/PhysRevA.39.2000} {\bibfield
  {journal} {\bibinfo  {journal} {Phys. Rev. A}\ }\textbf {\bibinfo {volume}
  {39}},\ \bibinfo {pages} {2000} (\bibinfo {year} {1989})}\BibitemShut
  {NoStop}%
\bibitem [{\citenamefont {Jonathan}\ \emph {et~al.}(1999)\citenamefont
  {Jonathan}, \citenamefont {Furuya},\ and\ \citenamefont
  {Vidiella-Barranco}}]{Jonathan1999}%
  \BibitemOpen
  \bibfield  {author} {\bibinfo {author} {\bibfnamefont {D.}~\bibnamefont
  {Jonathan}}, \bibinfo {author} {\bibfnamefont {K.}~\bibnamefont {Furuya}},\
  and\ \bibinfo {author} {\bibfnamefont {A.}~\bibnamefont
  {Vidiella-Barranco}},\ }\href {https://doi.org/10.1080/09500349908231366}
  {\bibfield  {journal} {\bibinfo  {journal} {J. Mod. Opt.}\ }\textbf {\bibinfo
  {volume} {46}},\ \bibinfo {pages} {1697} (\bibinfo {year}
  {1999})}\BibitemShut {NoStop}%
\end{thebibliography}
%

\end{document}